\documentclass[10pt,twocolumn,aps,english,pra,superscriptaddress,longbibliography]{revtex4-2} 
\usepackage{tikz} 
\usepackage{amsmath,amsfonts,amsthm, mathtools}
\usepackage{placeins}
\usepackage{float}
\usepackage{dsfont}
\usepackage{csquotes}
\usepackage{amssymb}
\usepackage{verbatim}
\usepackage{bm}
\usepackage{amsmath}
\usepackage{amssymb}
\usepackage{stmaryrd}
\usepackage{amsthm}
\usepackage[caption=false]{subfig}
\usepackage{physics}
\usepackage{bm}  

\newcommand{\yujie}[1]{\textcolor{orange}{\emph{[Yujie: #1]}}}

\newtheorem{proposition}{Proposition}

\makeatletter
\@ifundefined{textcolor}{}
{
 \definecolor{BLACK}{gray}{0}
 \definecolor{WHITE}{gray}{1}
 \definecolor{RED}{rgb}{1,0,0}
 \definecolor{GREEN}{rgb}{0,1,0}
 \definecolor{BLUE}{rgb}{0,0,1}
 \definecolor{CYAN}{cmyk}{1,0,0,0}
 \definecolor{MAGENTA}{cmyk}{0,1,0,0}
 \definecolor{YELLOW}{cmyk}{0,0,1,0}
 }
\makeatother

\begin{document}

\title{Temporally localized quantum operations on continuous-wave thermal light
}

\author{Yunkai Wang}
\altaffiliation{Current affiliation: Perimeter Institute for Theoretical Physics, Waterloo, Ontario N2L 2Y5, Canada; Institute for Quantum Computing and Department of Applied Mathematics, University of Waterloo, Waterloo, ON N2L 3G1, Canada}
\affiliation{Department of Physics, University of Illinois Urbana-Champaign, Urbana, Illinois 61801, USA}
\affiliation{Illinois Quantum Information Science \& Technology Center (IQUIST), University of Illinois Urbana-Champaign, Urbana, Illinois 61801, USA}
\author{Yujie Zhang}
\altaffiliation{Current affiliation: Perimeter Institute for Theoretical Physics, Waterloo, Ontario N2L 2Y5, Canada; Institute for Quantum Computing and Department of Physics \& Astronomy, University of Waterloo, Waterloo, ON N2L 3G1, Canada}
\affiliation{Department of Physics, University of Illinois Urbana-Champaign, Urbana, Illinois 61801, USA}
\affiliation{Illinois Quantum Information Science \& Technology Center (IQUIST), University of Illinois Urbana-Champaign, Urbana, Illinois 61801, USA}
\author{Virginia O. Lorenz}
\affiliation{Department of Physics, University of Illinois Urbana-Champaign, Urbana, Illinois 61801, USA}
\affiliation{Illinois Quantum Information Science \& Technology Center (IQUIST), University of Illinois Urbana-Champaign, Urbana, Illinois 61801, USA}
\date{\today }

\begin{abstract}
Previous work showed that thermal light with a blackbody spectrum cannot be decomposed into a mixture of independent localized pulses. However, we find that in the weak-source limit and under the assumption of a flat spectrum, the first non-vacuum term in the state expansion does form a mixture of such pulses. This decomposition is essential for quantum-enhanced astronomical interferometry, which typically operates on localized pulses even though stellar light is inherently continuous-wave. We present a quantum derivation of the van Cittert–Zernike theorem that incorporates finite bandwidth, thereby justifying the operations on localized pulses while processing continuous-wave thermal light. For general spectra in the weak-source limit, we establish a criterion under which correlations between pulses can be safely neglected. When this criterion is not met, we provide a corrected strategy that accurately accounts for both the spectral profile and the detector-defined pulse shape.
\end{abstract}

\maketitle


\textit{Introduction} - Thermal light, such as sunlight, is ubiquitous in nature and features in some protocols of quantum optics. Here, we use the term `thermal light' to refer to radiation emitted by a system in thermal equilibrium at a given temperature and exhibiting thermal photon statistics \cite{mandel1995optical}. Recent proposals have explored the use of distributed entanglement for astronomical interferometry to image stellar thermal sources \cite{gottesman2012longer}, which is an important application of quantum networks \cite{duan2010colloquium,wei2022towards,sangouard2011quantum,azuma2023quantum}. Subsequent work has investigated various enhancements to such schemes: using quantum memories to reduce the entanglement requirements \cite{khabiboulline2019quantum,khabiboulline2019optical}, employing nonlinear controlled quantum gates to improve performance \cite{czupryniak2023optimal}, and analyzing the entanglement cost of different protocols \cite{czupryniak2022quantum}. Other studies have demonstrated the advantage of using multiple entangled states without the need for quantum repeaters \cite{marchese2023large}, proposed continuous-variable versions of entanglement-assisted interferometry \cite{huang2024limited,wang2023astronomical,purvis2024practical}, and established fundamental limits along with optimal protocols under superselection rule constraints \cite{zhang2025criteria}.
It is commonly assumed that stellar light is in a thermal state upon detection on Earth's surface. In such schemes, it is common to interfere stellar light with terrestrial photons, the latter typically prepared in localized temporal modes. This raises a subtle issue: when a localized terrestrial photon is interfered with continuous-wave stellar light, how should one account for the mismatch in mode structure, and what are the consequences?




This question is related to the decomposition of thermal light into localized pulses, which has been discussed in a series of papers~\cite{chenu2015first,chenu2015thermal,branczyk2017thermal}. These works show that thermal light with a blackbody spectrum cannot be decomposed into a mixture of independent localized pulses. We will refer to any optical state associated with a localized temporal mode as a pulse and use the terms ``temporal mode'' and ``pulse'' interchangeably throughout this work.  
References \cite{chenu2015first,chenu2015thermal} demonstrate the impossibility of such a decomposition by showing that it leads to contradictions in the first- and second-order correlation functions. Attempts to resolve these contradictions using unphysical constructs, such as a trace-improper density matrix, still fail to reproduce both correlation functions accurately.
These results call into question the validity of typical analyses in quantum-enhanced astronomical interferometry that apply quantum operations independently to each localized temporal mode of thermal light without accounting for inter-temporal mode correlations \cite{khabiboulline2019quantum,khabiboulline2019optical,gottesman2012longer,czupryniak2023optimal,czupryniak2022quantum,marchese2023large,huang2024limited,wang2023astronomical,purvis2024practical,zhang2025criteria}. In this letter, we examine whether and when these temporal correlations have any observable consequences for such protocols.

We begin by considering a simple one-dimensional model  with thermal light propagating along a single spatial dimension, e.g., in optical fiber, and focusing on specific cases different from the blackbody spectrum studied in Refs.~\cite{chenu2015first,chenu2015thermal,branczyk2017thermal}. Specifically, we show that when the spectrum of thermal light is flat, the light can indeed be decomposed into a direct product of independent localized pulses. We show that this decomposition gives the correct first- and second-order correlation functions. We further find that in the weak strength limit, the leading non-vacuum term is a mixture of independent localized pulses.  We also observe that the trace-improper density matrix constructed in Ref.~\cite{chenu2015first,chenu2015thermal}, which yields the correct first-order correlation function, naturally emerges in a similar form as the leading non-vacuum term in the weak-strength limit. The additional terms in the decomposition ensure proper normalization of the density matrix, making the overall state physical.

We then analyze the impact of temporal mode correlations on quantum-enhanced astronomical interferometry \cite{khabiboulline2019quantum,khabiboulline2019optical,gottesman2012longer,czupryniak2023optimal,czupryniak2022quantum,marchese2023large,huang2024limited,wang2023astronomical,purvis2024practical,zhang2025criteria}. We present a complete derivation of the van Cittert–Zernike theorem within the quantum description of multimode thermal light with finite spectral bandwidth, followed by a discussion of how inter-mode correlations influence the measurement process in such schemes. For thermal light with a flat spectrum, operations can be performed in any set of localized temporal modes without the need to account for correlations between different modes. Furthermore, we consider the weak-source limit with a general spectrum and establish a criterion under which correlations between localized temporal modes can be neglected. When this criterion is not satisfied, we provide a strategy that accounts for these correlations for arbitrary spectral shapes and temporal mode profiles, tailored for experimental convenience. This addresses some of the naivety present in typical single-mode quantum optics analyses applied to quantum-enhanced astronomical interferometry. 
This framework also allows us to investigate the rich interplay among key properties of thermal light, including spatial correlations between the two telescopes, temporal correlations between localized modes induced by a non-flat spectrum, and the quantum statistical features of thermal light such as the Hanbury Brown–Twiss (HBT) effect \cite{glauber1963quantum,brown1956correlation,hanbury1979test}.

\textit{Preliminary} -
We first review the theoretical framework needed for the decomposition of a thermal state in the frequency domain into localized modes in the spatial domain. 
For simplicity, we first consider the one-dimensional case such that the wavevector $k$ is scalar.  We consider thermal light with a continuous spectrum consisting of $N\rightarrow\infty$ frequency modes $a_{m=1,2,\cdots N}$, and denote the wavevector corresponding to the $m$th mode by  $k_m$. After it has passed through a band-pass filter, which is commonly employed in astronomical interferometry \cite{monnier2003optical}, the light can be described as 
\begin{equation}\label{thermal_state}
\begin{aligned}
&\rho=\int %
\frac{ d^{2N}\vec{\alpha}}{\det(\pi\Gamma)}F(\vec{\alpha})\ket{\vec{\alpha}}\bra{\vec{\alpha}},\\
&F(\vec{\alpha})=\exp(-\vec{\alpha}^\dagger\Gamma^{-1}\vec{\alpha}),\quad \Gamma_{mn}=\delta_{mn}n_m,\\
&\\
\end{aligned}
\end{equation}
where $\ket{\vec{\alpha}}=\exp\left(\sum_m\alpha_m a_m^\dagger-\alpha_m^*a_m\right)\ket{0}$ is a multimode coherent state spanning all frequency modes. In this P-representation of the state \cite{mandel1995optical}, $\Gamma$ is the diagonal matrix of mean thermal photon numbers, with $n_m$ denoting the mean photon number of the $m$th mode. We assume $n_m\neq 0$ only when $k_m\in[k_0-\Delta k/2,k_0+\Delta k/2]$, corresponding to the photons transmitted by the band-pass filter centered at $k_0$ with bandwidth $\Delta k$. We will see that the spectrum described by $n_m$ is important for the decomposition of thermal light.
Using a construction similar to Refs.~\cite{branczyk2017thermal,blow1990continuum,rohde2007spectral}, we define the localized temporal modes as
$c_s = \sum_{m=1}^N O_{sm} a_m$, for $s = 1, 2, \dots, N$,
where each index $s$ labels a temporal mode $c_s$. The matrix $O$ can be chosen such that the resulting modes $c_s$ are both orthonormal and spatially localized.
This allows us to write Eq.~\ref{thermal_state} as 
\begin{equation}\label{eq:general_rho}
\begin{aligned}
&\rho=\int \frac{ \prod_{s=1}^N d^2\gamma_s}{\det(\pi\Lambda)}F(\vec{\gamma})\ket{\vec{\gamma}}\bra{\vec{\gamma}},\\
&F(\vec{\gamma})=\exp(-\vec{\gamma}^\dagger\Lambda^{-1}\vec{\gamma}),\quad \Lambda=(O^{-1})^\dagger \Gamma O^{-1},
\end{aligned}
\end{equation}
where $\ket{\vec{\gamma}}=\exp\left(\sum_s\gamma_s c_s^\dagger-\gamma_s^*c_s\right)\ket{0}$.
This is the decomposition derived in Ref. \cite{branczyk2017thermal}. 
For more details on the preliminary, see Sec. A of the Supplemental Material.

\textit{Decomposition of a thermal state into  localized pulses} - We now discuss the general decomposition of states in the one-dimensional case and address the relation of our discussion with the previous work of Ref. \cite{chenu2015first,chenu2015thermal,branczyk2017thermal}. 
We first show that, in general, thermal light cannot be decomposed into independent localized pulses.
We construct a set of orthonormal functions $\{ \omega_s(z) \}$ to serve as the pulse shapes, or equivalently their Fourier transforms ${\tilde{\omega}_s(k)}$, which must also form an orthonormal set
\begin{equation}\label{eq:omega1}
\int_{k_0-\Delta k/2}^{k_0+\Delta k/2}dk \tilde{\omega}_{s_1}^*(k)\tilde{\omega}_{s_2}(k)=\delta_{s_1s_2}.
\end{equation}
As an example, the wavefunction for a localized mode could be $\omega_s(z)\propto e^{ik_0(z-x_s)}\text{sinc}\frac{\Delta k(z-x_s)}{2}$, where $z$ denotes the spatial coordinate along the propagation direction. The pulses are separated by $2\pi / \Delta k$, and each has a spatial width on the order of $\Delta k^{-1}$, determined by the $k$-space bandwidth $\Delta k$. For more details on these calculations, see Sec. A of the Supplemental Material. If we calculate the  first-order correlation function between the two localized spatial modes $a_{s_1}$ and $a_{s_2}$
\begin{equation}\begin{aligned}\label{eq:omega2}
&G^{(1)}(s_1,s_2)
=
\int_{k_0-\Delta k/2}^{k_0+\Delta k/2}dk \tilde{\omega}_{s_1}^*(k)\tilde{\omega}_{s_2}(k) n(k)=0, 
\end{aligned}
\end{equation}
where $s_1\neq s_2  $.
Satisfying both Eq. \ref{eq:omega1} and Eq. \ref{eq:omega2} is equivalent to constructing a set of functions that are orthogonal with respect to two different weight functions, 
which is generally not possible for an arbitrary spectrum  $n(k)$, where we use $n_m = n(k_m)$ for notational convenience. In this work, we refer to a nonvanishing correlation function between different temporal modes as the presence of temporal-mode correlations. The modes are considered independent when no such correlations are present.

We now consider several special cases where such a decomposition either exists exactly or holds approximately. In studies of quantum-enhanced astronomical imaging at optical wavelengths \cite{khabiboulline2019quantum,khabiboulline2019optical,gottesman2012longer}, it is typically appropriate to assume the weak-source limit,  due to the extremely low mean photon number received in each localized temporal mode. In this weak limit, the decomposition of thermal light with a general spectrum, as given in Eq.~\ref{thermal_local}, can be expanded as follows:
\begin{equation}\begin{aligned}
&\rho\approx\ket{0}\bra{0}+\sum_{s,s'=1}^N\Lambda_{s's}\ket{s'}\bra{s}+\cdots,
\end{aligned}
\end{equation}
where $\ket{s} = c_s^\dagger \ket{0}$ and we assume weak source limit $|\lambda_{\text{max}}(\Lambda)| \to 0$. Note that, in general, $\Lambda$ is not diagonal,  so neighboring  modes   $c_s$ and $c_{s+1}$ may exhibit correlations.
This means for the case of a general spectrum, even in the weak limit, we cannot directly assume the state is a mixture of independent localized pulses.

Consider the decomposition of thermal light with a flat spectrum by
choosing $n_m=n$ for $\forall m$. This choice results in $\Lambda_{ss'}=\delta_{ss'}\Lambda_{ss}$, which means the different $c_s$ modes are independent of each other and Eq.~\ref{eq:general_rho} is simplified as 
\begin{equation}
\begin{aligned}\label{thermal_local}
&\rho=\otimes_s\rho'_s,\,\,\rho'_s
=\int \frac{d^2\gamma_s}{\pi n}\exp(-|\gamma_s|^2/n)\ket{\gamma_s}\bra{\gamma_s}.
\end{aligned}
\end{equation}
Note that each localized mode $c_s$ is in a thermal state $\rho_s'$ and all modes $c_s$ are independent of each other.

By assuming both a flat spectrum and the weak source limit, we find the decomposition 
\begin{equation}
\begin{aligned}\label{thermal_weak}
&\rho\approx\ket{0}\bra{0}+\epsilon\sum_{s=1}^N\ket{s}\bra{s}+\cdots,
\end{aligned}
\end{equation}
where $\Lambda_{ss}=\epsilon$ and the $O(\epsilon)$ term represents the first non-vacuum contribution and corresponds to a mixture of independent localized pulses, where all pulses are in the vacuum state except for one that contains a single photon. We do not know which pulse contains the single photon, as reflected by the mixed state. This is a very common intuitive picture used in analyses of 
astronomical interferometry enhanced by quantum technologies \cite{khabiboulline2019quantum,khabiboulline2019optical,gottesman2012longer,czupryniak2023optimal,czupryniak2022quantum,marchese2023large,huang2024limited,wang2023astronomical,purvis2024practical,zhang2025criteria}. We validate our decomposition by showing it reproduces the expected thermal-state correlation functions without requiring an unphysical trace-improper density matrix whose trace exceeds one, unlike previous discussions \cite{chenu2015thermal,chenu2015first}, and naturally maintains trace normalization as the volume increases. For further discussion of the correlation functions, their relation to previous works, and specific cases of deviation from a flat spectrum, please refer to Sec. B of the Supplemental Material.

\textit{Astronomical interferometric imaging of a thermal source with a general spectrum} - We now consider astronomical interferometry as in Ref.~\cite{gottesman2012longer}, which typically interferes continuous-wave stellar light with a terrestrial photon prepared in a localized temporal mode. As shown earlier, a flat spectrum allows thermal light to be decomposed into a product of independent, localized pulses in orthonormal temporal modes, and the interference between terrestrial and stellar light can then be safely treated as independent across modes. However, with a non-flat spectrum, ignoring inter-mode correlations can introduce inaccuracies. We now establish criteria under which such correlations can be neglected and present a corrected estimation procedure when the criteria is not satisfied. These corrections can be applied in post-processing, enabling flexible mode choices convenient for experimental implementation while maintaining accuracy.


To quantify the influence of a general spectrum of thermal light on interferometric imaging, we rederive the van Cittert–Zernike theorem within a multimode quantum optics framework.
We start from the state of a general incoherent source \cite{mandel1995optical} 
\begin{equation}
\begin{aligned}
&\rho=\int d^{2MW}\vec{\alpha}\,\Phi(\vec{\alpha})\ket{\vec{\alpha}}\bra{\vec{\alpha}},\,\, \Phi(\vec{\alpha})=\frac{\exp(-\vec{\alpha}^\dagger\Gamma^{-1}\vec{\alpha})}{\det(\pi\Gamma)},\\
\end{aligned}
\end{equation}
where $\Gamma_{i_1m_1,i_2m_2}=\delta_{i_1i_2}\delta_{m_1m_2}n_{i_1m_1}$, the multimode coherent state  $\ket{\vec{\alpha}}=\otimes_{i=1}^W\otimes_{m=1}^M\exp(\alpha_{im}\hat{a}_{i,k_m}^\dagger-\alpha_{im}^*\hat{a}_{i,k_m})\ket{0}$  is used as a basis to represent the thermal state,  the index $i = 1, 2, \ldots, W$ labels spatial points on the source, and we consider the continuous spatial intensity distribution in the limit $W \to \infty$. The index $m = 1, 2, \ldots, M$ labels  frequencies $k_m$, with the continuous frequency spectrum recovered in the limit $M \to \infty$. The quantity $n_{im}$ denotes the mean photon number emitted at frequency $k_m$ from the $i$th spatial point on the source. For simplicity, we model the source as one-dimensional by neglecting variations along the $y$-axis. Thus, the spatial coordinates of the $i$th source point are $(x_i, d)$, where $x_i$ is its position along the $x$-axis and $d$ is the  distance between the source and the telescope array along the $z$-axis.

We consider the case of two telescopes. Given that the astronomical source is in the far-field limit relative to the telescopes on Earth, we assume the incoming light propagates exclusively along the $z$-axis near the telescopes. Let the two lenses be located at positions $(u_0, 0)$ and $(u_1, 0)$. Under this assumption, the source annihilation operators $\hat{a}_{i,k_m}$ evolve as 
$\hat{a}_{i,k_m}\rightarrow \int dz \sqrt{\eta}e^{i\phi(k_m,u_1,z,x_i)}\hat{b}_{z0}
+\int dz \sqrt{\eta}e^{i\phi(k_m,u_2,z,x_i)}\hat{b}_{z1}+\hat{v}_{i,k_m}$,    
where $\hat{v}_{i,k_m}$ describes the environmental modes and $\eta$ quantifies the transmission efficiency from the source to the spatial modes near the telescopes. The operators $\hat{b}_{z0}$ and $\hat{b}_{z1}$ correspond to the spatial modes immediately above the first and second telescopes, located at positions $(u_0, z)$ and $(u_1, z)$, respectively. The accumulated propagation phase is given by
$\phi(k,u,z,x)=k\sqrt{(x-u)^2+(d-z)^2}$.
With this setup, we can now determine the quantum state of the light field near the two telescopes. In the weak-source limit, we expand the total state into Fock states in each localized temporal mode:
\begin{equation}\label{eq:rho_expansion_stellar}
\begin{aligned}
&\rho
=\ket{0}\bra{0}+\sum_{s,s'=1}^N\sum_{p,q\in\{0,1\}}\vec{\omega}_{s' p}^\dagger \Lambda \vec{\omega}_{sq}\ket{s',p}\bra{s,q}+\cdots\\
&\Lambda_{lt,pq}=\sum_{i=1}^W \sum_{m=1}^M \eta \, n_{im}e^{i(\phi(k_m,u_p,z_q,x_i)-\phi(k_m,u_l,z_t,x_i))},
\end{aligned}
\end{equation}
where $\ket{s,p}=\hat{c}_{sp}^\dagger\ket{0}$,  $\hat{c}_{sp}^\dagger=\sum_{n=1}^N\omega_{s}(z_n)b_{z_np}^\dagger\ket{0}$, we define the vector $\vec{\omega}_{sq}$ such that the double sum over spatial coordinates is expressed as matrix multiplication $\sum_{n,m}\omega_{s'}^*(z_{n}) \Lambda_{p n,qm}\omega_{s}(z_{m})=\vec{\omega}_{s' p}^\dagger \Lambda\vec{\omega}_{sq}$, $\omega_{s}(z)$ is the $s$th localized temporal mode, we denote the position of the $n$th point near the two telescopes as $(u_{0,1}, z_n)$, and we take the limit $N \to \infty$ to describe a continuous light field. Note that, in general, nearby temporal modes exhibit correlations $\vec{\omega}_{s+1, p}^\dagger \Lambda \vec{\omega}_{sq}\neq 0$, so we cannot claim that the state is a mixture of independent localized pulses. 


We now investigate the impact of a non-flat spectrum on the estimation results. For the measurement applied to the $s$th temporal mode, the state is projected onto $\ket{00}_s$ when no photon is detected, and onto the state $(\ket{01}_s \pm e^{i\theta} \ket{10}_s)/\sqrt{2}$ when a single photon is detected. Here,
$\ket{01}_s = \sum_{n=1}^N \omega_s(z_n) \hat{b}_{0 z_n}^\dagger \ket{0}$ and
$\ket{10}_s = \sum_{n=1}^N \omega_s(z_n) \hat{b}_{1 z_n}^\dagger \ket{0}$.
Suppose we perform this measurement successively across $N$ temporal modes in time and detect a total of $Q$ photons. Let $\vec{s} = [s_1, s_2, \ldots, s_Q]$ denote the indices of the temporal modes in which the photons are detected. We define $\vec{\delta} = \{e^{i\theta}, -e^{i\theta}\}^{\otimes Q}$ to label the possible outcomes of detecting $Q$ photons, which projects the state onto 
\begin{equation}\begin{aligned}\label{eq:P_full_main}
&\ket{\vec{s},\vec{\delta}}=\sum_{\vec{p}\in\{0,1\}^{\otimes Q}}\left(\prod_{i=1}^Q\delta_i^{p_i}\right)\ket{\vec{s},\vec{p}}/2^{Q/2},\\
&P(\vec{s},\vec{\delta})=\sum_{\vec{p},\vec{q}}\sum_\mathbb{P}\left[\prod_{i=1}^Q\frac{1}{2} \delta_i^{p_i+q_i}\,\,\vec{\omega}_{s_i p_i}^\dagger\Lambda\vec{\omega}_{s_{\mathbb{P}_i}q_{\mathbb{P}_i}}\right],\\
\end{aligned}
\end{equation} 
where $\sum_\mathbb{P}$ means the sum over permutations, $\ket{\vec{s},\vec{p}}=\otimes_{i=1}^Q(\sum_{n=1}^N\omega_{s_i}(z_n)b_{p_iz_n}^\dagger)\ket{0}$.

The imaging task aims to estimate the intensity distribution $n_{im}$ encoded in $\Lambda$, by inferring it from the empirical distribution obtained by sampling the true probability distribution in Eq. \ref{eq:P_full_main}. Correlations between different temporal modes, i.e. $\vec{\omega}_{s' p}^\dagger \Lambda\vec{\omega}_{sq}\neq 0$ for $s\neq s'$, can influence the probability $P(\vec{s}, \vec{\delta})$ and thus affect astronomical imaging. However, unlike Eq. \ref{eq:rho_expansion_stellar}, which involves all modes, $P(\vec{s}, \vec{\delta})$ depends only on correlations between modes where photons are actually detected. For light with bandwidth $\Delta k$, these correlations decay over a timescale $\sim \Delta k^{-1}$; when detection events are separated by $\Delta t \gg \Delta k^{-1}$, such correlations can be neglected. This intuition is formalized below to establish the criteria under which inter-mode correlations can be safely ignored.

\begin{proposition}\label{prop 1}
Assuming $n_{im} = I_i f_m$,  where $f_m$ is the normalized spectrum satisfying $\sum_m f_m = 1$, implying that the relative intensity $I_i$ across spatial points is independent of frequency—reasonable within a narrow bandwidth $\Delta k \ll k_0$—and $\omega_r(z) = \omega(z - z_r)$, and further approximating the phase $\phi(k, u, z, x)$ by its value at $k = k_0$, thereby neglecting its frequency dependence, we compute the deviation between the actual probability distribution $P(\vec{s},\vec{\delta})$—which includes correlations between different temporal modes under a general spectrum—and the distribution $P_0(\vec{s},\vec{\delta})$ assuming uncorrelated temporal modes:
\begin{equation}
\begin{aligned}\label{eq:deviation_P}
|P(\vec{s},\vec{\delta})-P_0(\vec{s},\vec{\delta})|=(2\epsilon)^Q\sum_{\mathbb{P}\neq I}\prod_{i=1}^Q\xi(x_{s_i}-x_{s_{\mathbb{P}_i}} ),
\end{aligned}
\end{equation}
where $\sum_{\mathbb{P}\neq I}$ means the sum over permutations except the identity, $\epsilon =\sum_{i=1}^W\eta I_i$, and the function quantifying correlations $\xi(x_{s_1}-x_{s_2})=\Omega_{s_1s_2}/\Omega_{ss}$, we find $\xi(\Delta x)\rightarrow0$ as $|\Delta x|\gg2\|P(k)\|_{\infty}+\Delta k\|P'(k)\|_{\infty}$, $P(k)={ f(k)|\tilde{\omega}(k)|^2}/\left({\int_{k_0-\Delta k/2}^{k_0+\Delta k/2}dk f(k)|\tilde{\omega}(k)|^2}\right)$, $P_0(\vec{s},\vec{\delta})=\prod_{i=1}^Q\left(\frac{1}{2}\sum_{p,q\in\{0,1\}}\delta_i^{p+q}\vec{\omega}_{s_ip}^\dagger\Lambda\vec{\omega}_{s_iq}\right)$, $\Omega_{s_1s_2}=\sum_{m=1}^Mf_m\tilde{\omega}^*_{s_1}(k_m)\tilde{\omega}_{s_2}(k_m)$, $P'(k)=dP(k)/dk$.
\end{proposition}



The factor $\Omega_{s_1 s_2}\propto \vec{\omega}_{s_1 p}^\dagger \Lambda\vec{\omega}_{s_2q}$ captures the effect of correlations between different temporal modes. Clearly, when $f_m = f$ is constant for all $m$, we have $\Omega_{s_1 s_2} = \delta_{s_1 s_2} f$, and no inter-mode correlation arises, hence $P(\vec{s},\vec{\delta})=P_0(\vec{s},\vec{\delta})$. However, for a general spectrum, $\Omega_{s_1 s_2}\neq 0$ introduces additional contributions to the probability distribution, hence $P(\vec{s},\vec{\delta})\neq P_0(\vec{s},\vec{\delta})$. We  take $\omega_r(z) = \omega(z - z_r)$, corresponding to a set of localized, orthonormal temporal modes that are translations of one another—a typical choice in experimental implementations.  Although $\phi(k, u, z, x)$ depends on $k$, we approximate it by its value at $k_0$; but frequency dependence can be included via a Taylor expansion of $\phi(k, u, z, x)$ around $k_0$ if needed.


Clearly, $P_0(\vec{s}, \vec{\delta})$ is of order $\epsilon^Q$, as $\epsilon$ corresponds to the mean photon number per mode—after accounting for losses from the stellar source to telescopes—while $P_0(\vec{s}, \vec{\delta})$ denotes the probability of detecting $Q$ photons in the temporal modes indexed by $\vec{s}$. Note that the summation excludes the identity permutation, so as long as the detected photons are sufficiently separated in time—i.e., $|\Delta x|\gg2\|P(k)\|_{\infty}+\Delta k\|P'(k)\|_{\infty}$ for all pairs $s_i, s_j$—the difference $|P(\vec{s}, \vec{\delta}) - P_0(\vec{s}, \vec{\delta})|$ becomes negligible. In practice, $\vec{s}$ and $\vec{\delta}$ correspond to the detection outcomes from measuring $N$ localized temporal modes and observing $Q$ photons. 
$\xi(\Delta x)$ can be computed in practice by characterizing the spectral profile $f(k)$ through preliminary measurements that record the intensity at each frequency. 
 This allows Eq.~\ref{eq:deviation_P} to serve as an estimate for whether further correction is necessary. If a significant fraction of photons are detected in closely spaced temporal modes such that $\sum_{\mathbb{P} \neq I} \prod_{i=1}^Q \xi(x_{s_i} - x_{s_{\mathbb{P}_i}})$ cannot be assumed to be negligible, then data postprocessing must account for the resulting correlations, as detailed in the following proposition.

\begin{proposition}\label{prop 2}

Assuming $n_{im} = I_i f_m$, $\sum_m f_m=1$ and $\omega_r(z) = \omega(z - z_r)$, and further approximating the phase $\phi(k, u, z, x)$ by its value at $k = k_0$, the actual probability distribution—accounting for correlations between different temporal modes under a general spectrum—is given by
\begin{equation}
\begin{aligned}
&P(\vec{s},\vec{\delta})=\frac{\epsilon^Q}{2^Q}\sum_{\mathbb{P}}\prod_{i=1}^Q[(\delta_ig^*+\delta_{\mathbb{P}_i}g+1+\delta_i\delta_{\mathbb{P}_i})\Omega_{s_i,s_{\mathbb{P}_i}}],
\end{aligned}
\end{equation}
where $\epsilon g=\sum_{i=1}^W\eta I_i\exp\left(ik_0\frac{u_1^2-u_0^2}{2d}+ik_0\frac{x_i(u_0-u_1)}{d}\right)$.
\end{proposition}

In interferometric imaging, the goal is to estimate the coherence function $g$, which is a function of the intensity distribution $n_{im}$. Given the probability distribution $P(\vec{s}, \vec{\delta})$, which explicitly accounts for temporal-mode correlations, the parameter $g = |g| e^{i\theta}$ can be estimated using a standard maximum likelihood approach by solving $\frac{\partial{P(\vec{s},\vec{\delta})}}{\partial |g|}\bigg|_{|g|=\hat{|g|}}=\frac{\partial{P(\vec{s},\vec{\delta})}}{\partial \theta}\bigg|_{\theta=\hat{\theta}}=0$.

As an example, we consider a spectrum of the form $f_m \propto \exp\left( -\frac{(k_m - k_0)^2}{\Delta k^2} \right)$ and a pulse shape $\tilde{\omega}(k_m) \propto \exp\left( -\frac{(k_m - k_0)^2}{2\Delta k^2} \right)$ for $k_m \in [k_0 - \Delta k/2, k_0 + \Delta k/2]$. This pulse shape can be approximately realized by applying a Gaussian-shaped bandpass filter in the frequency domain and detecting within a time window of duration roughly $1/\Delta k$.  In the case where two photons are detected at positions $x_{s_1}$ and $x_{s_2}$, Proposition \ref{prop 1} provides the bound 
\begin{equation}
\begin{aligned}
&|P(\vec{s},\vec{\delta})-P_0(\vec{s},\vec{\delta})|\leq(2\epsilon)^2|\xi(\Delta x )|^2,\quad\xi(\Delta x)
\leq \frac{4.93}{|\Delta x|\Delta k},
\end{aligned}
\end{equation}
where we define the distance between the two detected photons $\Delta x=x_{s_1}-x_{s_2}$.
If $\Delta x \sim 1/\Delta k$, the correction provided by Proposition \ref{prop 2} becomes necessary. We consider the example where the detection outcome is $\vec{\delta} = [1, -1]$, which corresponds to a projection onto the state $(\ket{01}_{s_1}+\ket{10}_{s_1})\otimes(\ket{01}_{s_2}+\ket{10}_{s_2})/2$. The probability is given by
\begin{equation}
\begin{aligned}
&P(\vec{s},\vec{\delta})=\epsilon^2(1-|g|^2\cos^2\theta\,\Omega_{ss}^2+|g|^2\sin^2\theta\,\Omega_{s_1s_2}\Omega_{s_2s_1}),\\
\end{aligned}
\end{equation}
where $\Omega_{s_1s_2}=C(\Delta x)\frac{\exp\left(-\frac{\Delta k^2\Delta x^2}{8}\right)}{2\sqrt{2\pi}(\text{erf}(1/2))^2}$, $C(\Delta x)=\text{erf}\left[\frac{2-i\Delta x\Delta k}{2\sqrt{2}}\right]+\text{erf}\left[\frac{2+i\Delta x\Delta k}{2\sqrt{2}}\right]$. $P(\vec{s},\vec{\delta})$ is the accurate formula that accounts for the correlation and can be used to estimate $g$ using methods such as maximum likelihood estimation.

We emphasize that both Proposition 1 and 2 rely on the weak-source limit, meaning that most temporal modes contain only vacuum, and correlations between temporal modes need not be taken into account in many cases. 
As the source becomes stronger, single-photon detection at each temporal mode is no longer an appropriate measurement strategy; in such cases, local heterodyne detection may be more suitable. 
Our framework also allows us to explore the Hanbury Brown–Twiss (HBT) effect and the operation of intensity interferometers under general spectral conditions and finite bandwidth. 
As a consistency check, we verify that the model reproduces the quantum statistical correlations expected for thermal light. For a flat spectrum, successful operation of the intensity interferometer requires simultaneous photon detection at both telescopes. However, in the case of a non-flat spectrum, we show that even when the photons are detected at different times, it is still possible to extract information about $g$.
A discussion of stronger sources imaged via local heterodyne detection, the HBT effect, more details of the illustrative examples, and the proofs of the two propositions are provided in Sec. C of the Supplemental Material.

\textit{Conclusion} - We have shown that thermal light with a flat spectrum is a direct product of independent localized pulses. In the weak-light limit, the first non-vacuum term of the expansion of thermal light with a flat spectrum in the Fock basis is a mixture of independent localized pulses. Our work provides a quantum derivation of the van Cittert–Zernike theorem that incorporates finite spectral bandwidth and a multimode description of thermal light. Building on this framework, we examined the effect of a general spectrum when an astronomical interferometer measures continuous-wave thermal light using localized temporal modes. We established criteria under which inter-mode correlations can be neglected, and proposed a corrected estimation strategy for cases where these correlations are significant. Our results serve as a prerequisite for quantum protocols involving thermal light emitted by naturally occurring sources, enabling the use of localized temporal modes that are convenient for experimental implementation. We expect this theoretical framework to be readily generalizable to other quantum protocols that involve operations on thermal light emitted by natural sources.
Moreover, our formulation captures the interplay between spatial correlations across telescopes, temporal correlations from spectral structure, and the quantum statistical properties of thermal light—revealing a rich and fundamentally interesting coupling between different aspects of the light.

\textit{Data availability} -  All codes used  in this paper are available at: https://github.com/ykwang-phys/thermal-light


\section*{Acknowledgements}

We are especially grateful to John E. Sipe for valuable feedback and comments. We also thank  Paul Kwiat, Eric Chitambar, Andrew Jordan, John D. Monnier, Shayan Mookherjea, Michael G. Raymer, and Brian J. Smith for helpful discussions. This work was supported by the multiuniversity National Science Foundation Grant No. 1936321 and No. 2326803.




\appendix

\onecolumngrid

\section{Preliminary}\label{SI:preliminary}

We here provide further details on the theoretical framework required to decompose a thermal state in the frequency domain into localized modes in the spatial domain. 
As derived in 
Ref.~\cite{mandel1995optical}, the density matrix of blackbody radiation 
\begin{equation}
\begin{aligned}
&\rho=\otimes_{\vec{k},\chi}\rho_{\vec{k},\chi},\\
&\rho_{\vec{k},\chi}=\sum_{p}\frac{n_{\vec{k},\chi}^p}{(n_{\vec{k},\chi}+1)^{p+1}}\ket{p}\bra{p},\\
&n_{\vec{k},\chi}=1/[\exp(\hbar\omega_{\vec{k}}/k_B T)-1],
\end{aligned}
\end{equation}
where $\rho_{\vec{k},\chi}$ is a thermal state defined in the frequency modes $a_{\vec{k},\chi}$  
 \cite{weedbrook2012gaussian}, which are labeled by the wavevector $\vec{k}$ and  the polarization $\chi$, satisfying $[a_{\vec{k}_1,\chi_1},a_{\vec{k}_2,\chi_2}^\dagger]=\delta_{\vec{k}_1\vec{k}_2}\delta_{\chi_1\chi_2}$, $\ket{p} = \frac{(\hat{a}_{\vec{k},\chi}^\dagger)^p}{\sqrt{p!}} \ket{0}$ denotes the $p$-photon Fock state in the frequency mode specified by the wavevector $\vec{k}$ and polarization $\chi$.
From this equation, it can be seen that blackbody radiation is the direct product of thermal states with different $\vec{k},\chi$ and that each of the frequency modes is independent. The spectrum of blackbody radiation is described by $n_{\vec{k},\chi}$ and follows a Bose-Einstein distribution. 

Filtering a thermal state in the frequency domain can be modeled as a beamsplitter transformation applied independently to each frequency mode: \( a_{\vec{k},\chi} \rightarrow \sqrt{\eta_{\vec{k},\chi}}\, a_{\vec{k},\chi} + \sqrt{1 - \eta_{\vec{k},\chi}}\, v_{\vec{k},\chi} \), where \( \eta_{\vec{k},\chi} \in [0,1] \) is the transmissivity of the filter, and \( v_{\vec{k},\chi} \) is an auxiliary vacuum mode \cite{leonhardt2003quantum,rohde2007spectral}. Since the original state is a product \( \rho = \otimes_{\vec{k},\chi} \rho_{\vec{k},\chi} \), where each \( \rho_{\vec{k},\chi} \) is a single-mode thermal state of the form \( \rho_{\vec{k},\chi} = \sum_{p=0}^\infty \frac{n_{\vec{k},\chi}^p}{(n_{\vec{k},\chi}+1)^{p+1}} \ket{p}\bra{p} \), the beamsplitter transformation acts independently on each mode and preserves the diagonal-in-Fock basis nature of the thermal state. Tracing out the vacuum environment mode \( v_{\vec{k},\chi} \) is equivalent to introducing loss, which attenuates the mean photon number to \( \eta_{\vec{k},\chi} n_{\vec{k},\chi} \) but does not introduce any coherence or off-diagonal elements. The output state in each mode remains a thermal state of the same form, with a modified spectrum. Thus, the overall filtered state is still a product of thermal states, \( \rho' = \otimes_{\vec{k},\chi} \rho'_{\vec{k},\chi} \), where \( \rho'_{\vec{k},\chi} \) has mean photon number \( \eta_{\vec{k},\chi} n_{\vec{k},\chi} \).

For simplicity, we consider the one-dimensional case where $\vec{k} = k$ is a scalar, and we fix the polarization to a single mode $\chi$. We introduce the subscript $m$ to denote frequency components $k_m$ for notation convenience, while still treating the thermal light as having a continuous spectrum  composed of $N \to \infty$ frequency modes $a_m$ with $m = 1, 2, \ldots, N$  and replace the sum with integral as needed. These modes have passed through a band-pass filter and are described as follows:
\begin{equation}\label{SI_eq:thermal_state}
\begin{aligned}
&\rho=\otimes_{s=1}^N\rho_s=\int \frac{ \prod_{s=1}^N d^2\alpha_s}{\det(\pi\Gamma)}F(\vec{\alpha})\ket{\vec{\alpha}}\bra{\vec{\alpha}},\quad F(\vec{\alpha})=\exp(-\vec{\alpha}^\dagger\Gamma^{-1}\vec{\alpha}),\\
&\ket{\vec{\alpha}}=\exp\left(\sum_s\alpha_s a_s^\dagger-\alpha_s^*a_s\right)\ket{0},\quad \Gamma_{mn}=\delta_{mn}n_m,
\end{aligned}
\end{equation} 
where $n_m$ is the mean photon number in the $m$th mode. We assume $n_m \neq 0$ only for frequencies within the passband, i.e., when $k_m \in [k_0 - \Delta k/2,, k_0 + \Delta k/2]$, where $\Delta k$ denotes the bandwidth of the filter. As we will see, the spectral profile specified by ${n_m}$ plays a crucial role in the decomposition of thermal light.

Using a construction similar to Ref.~\cite{branczyk2017thermal}, we define
\begin{equation}
c_s=\sum_{m=1}^NO_{sm}a_m,\quad s=1,2,\cdots N,
\end{equation}
where $O_{sm}$ defines the transformation used to construct a set of modes $c_s$, which can be made spatially localized and orthonormal. To ensure that the modes $c_s$ are orthonormal to one another, we require
\begin{equation}\begin{aligned}
&[c_{s_1},c_{s_2}^\dagger]=\sum_{m_1,m_2=1}^N O_{s_1m_1}O_{s_2m_2}[a_{k_{m_1}},a_{k_{m_2}}^\dagger]=[OO^\dagger]_{s_1s_2}=\delta_{s_1s_2}.    
\end{aligned}
\end{equation}
This allows us to write Eq.~\ref{SI_eq:thermal_state} as 
\begin{equation}\label{general_rho}
\begin{aligned}
&\rho=\int \frac{ \prod_{s=1}^M d^2\gamma_s}{\det(\pi\Lambda)}F(\vec{\gamma})\ket{\vec{\gamma}}\bra{\vec{\gamma}},\quad F(\vec{\gamma})=\exp(-\vec{\gamma}^\dagger\Lambda^{-1}\vec{\gamma}),\\
&\Lambda=C^\dagger \Gamma C,\quad C=O^{-1},\quad\ket{\vec{\gamma}}=\exp\left(\sum_s\gamma_s c_s^\dagger-\gamma_s^*c_s\right)\ket{0}.
\end{aligned}
\end{equation}
 This construction of modes as superpositions of frequency modes follows the interpretation presented in Rohde's work  \cite{rohde2007spectral}. While we adopt this framework, we further examine how thermal light with a general spectrum is measured using localized temporal modes. Note that for a general spectrum—including the blackbody spectrum—the spatially localized modes $c_s$ are not independent of one another and $\rho$ has correlations between different modes, consistent with the findings of Refs.~\cite{chenu2015first, chenu2015thermal}. 

Given the general formalism above for constructing spatially localized modes, we now present a specific example by choosing $O_{sm} = \frac{1}{\sqrt{N}} e^{i k_m x_s}$. When $s = 0$, this corresponds to the natural choice of an equal superposition of frequency modes, resulting in a localized wave packet, whereas $s \neq 0$ leads to a shift of the mode in spatial space. To ensure that the resulting set of localized pulses is orthonormal, we require
\begin{equation}\begin{aligned}
[OO^\dagger]_{s_1s_2}&=\sum_{m=1}^N\frac{1}{N}\exp(ik_m(x_{s_1}-x_{s_2}))=\int_{k_0-\Delta k/2}^{k_0+\Delta k/2}\frac{dk}{\Delta k/N}\frac{1}{N}\exp(ik(x_{s_1}-x_{s_2}))\\
&=e^{ik_0 x}\text{sinc}(\Delta k \Delta x/2 )=\delta_{s_1s_2},
\end{aligned}
\end{equation}
where $\text{sinc}(\Delta k \Delta x / 2) = \frac{\sin(\Delta k \Delta x / 2)}{\Delta k  \Delta x / 2}$ and $\Delta x = x_{s_1} - x_{s_2}$. In this expression, we have replaced the discrete sum with an integral. To ensure that the constructed modes form an orthonormal basis, we require $\Delta x = 2\pi n / \Delta k$ for some integer $n$. A natural choice is therefore $x_s = 2\pi s / \Delta k$. With this choice, the wavefunction of the state in frequency mode $a_m$ is given by
\begin{equation}
\phi_m(z)=e^{ik_mz}.
\end{equation}
So, the wavefunction of states in $c_s$ can be derived as $\omega_s(z)$ below:
\begin{equation}
\begin{aligned}
&a_m^\dagger\ket{0}=\int dz \phi_m(z) \ket{z},\\
&c_s^\dagger\ket{0}=\sum_m O_{sm}^*a_m^\dagger\ket{0}=\int dz\sum_{m=1}^N O_{sm}^* \phi_m(z)\ket{z}\\
&=\int dz\int_{k_0-\Delta k/2}^{k_0+\Delta k/2} \frac{dk}{\Delta k/N} \frac{e^{-ik x_s}}{\sqrt{N}}e^{ik z}\ket{z}=\int dz\omega_s(z)\ket{z},
\end{aligned}\end{equation}
\begin{equation}
\begin{aligned}
&\omega_s(z)=\int_{k_0-\Delta k/2}^{k_0-\Delta k/2} \frac{dk}{\Delta k}e^{ik(z-x_s)}\sqrt{N}=\sqrt{N}e^{ik_0(z-x_s)}\text{sinc}\frac{\Delta k(z-x_s)}{2}.    
\end{aligned}
\end{equation}
This demonstrates that the states in the $c_s$ modes are spatially localized, and one can readily verify that they are orthonormal. The spacing between neighboring pulses is $2\pi / \Delta k$, and the width of each pulse is on the order of $\Delta k^{-1}$. Notably, the bandwidth in frequency $k$-space, $\Delta k$, sets the characteristic width of the spatially localized pulses in this construction.


In practice, the measurement device projects the incoming field onto a set of localized temporal modes $\omega_s(z)$—for example by integrating over a finite detection window of duration $\delta z$ after band filter. The resulting transformation matrix $O$ is obtained by taking the Fourier transform
$\tilde\omega_s(k)=\int\omega_s(z)e^{-ikz}dz$
of each temporal mode $\omega_s(z)$. Because each $\omega_s(z)$ is confined to a time interval of width $\delta z$.  Its spectral variance $\delta k$ obeys the approximate uncertainty relation $\delta z \delta k \sim 1$, where $\delta k$ can be explicitly computed from the known shape of the detected temporal mode.
And by Chebyshev’s inequality
$\displaystyle\int_{|k-k_0|>c/\delta z}|\tilde\omega_s(k)|^2dk\lesssim 1/{c^2}$,
 at least a fraction $1-2/c^2$ of the spectral energy lies in $|k-k_0|\le c/\delta z$. Choosing $\Delta k \sim c/\delta z$ therefore captures the bulk of the spectrum, justifying the above formalism as a good approximation for practical applications involving temporally localized measurements.

\section{Decomposition for a simple 1D model}\label{SI:1D model}
\subsection{First- and second-order correlation functions for our decomposition}\label{Appendix:correlation}

In this subsection, we will explicitly calculate the  first- and second-order correlation functions for the decomposition of thermal states. First, let us recall the thermal state for the three cases, as described in the main text. The thermal state in the general case is given in Eq. \ref{SI_eq:thermal_state}.
We consider the decomposition of thermal light with a flat spectrum by setting $n_m = n$ for $\forall m$, resulting in the state:
\begin{equation}
\begin{aligned}\label{SI_eq:thermal_local}
&\rho=\otimes_s\rho'_s,\\
&\rho'_s=\int \frac{d^2\bar{\gamma}_s}{\pi}\exp(-|\bar{\gamma}_s|^2)\ket{\gamma_s}\bra{\gamma_s}=\int \frac{d^2\gamma_s}{\pi n}\exp(-|\gamma_s|^2/n)\ket{\gamma_s}\bra{\gamma_s}.
\end{aligned}
\end{equation}
If we further consider the weak limit, $n\rightarrow 0$, the decomposition in Eq.~\ref{SI_eq:thermal_local} can be expanded as 
\begin{equation}
\begin{aligned}\label{SI_eq:thermal_weak}
&\rho=\left(\frac{1}{1+n}\right)^N\bigg[\otimes_s\rho_s^{(0)}+\epsilon\sum_s\rho_1^{(0)}\otimes\rho_2^{(0)}\otimes\cdots\otimes\rho_s^{(1)}\otimes\cdots\otimes\rho_N^{(0)}+O(\epsilon^2)\bigg].
\end{aligned}
\end{equation}

For the general case described in the frequency domain in Eq. \ref{SI_eq:thermal_state}, we find
\begin{equation}
\begin{aligned}\label{general_G}
&G^{(1)}(z)=\sum_m {n}_m/L,\\
&G^{(2)}(z)=\sum_m 2{n}_m^2/L^2+\sum_{m_1\neq m_2}{n}_{m_1}{n}_{m_2}\bigg(\frac{1}{L^2}
+\frac{1}{L^2}e^{i(\kappa_{m_2}-\kappa_{m_1})z}\bigg),
\end{aligned}
\end{equation}
where $\kappa_m=2\pi m/L$, 
$G^{(1)}(z)$ is the shorthand notation of the first-order correlation function  $G^{(1)}(r=z,t=0;r=0,t=0)$, $G^{(2)}(z)$ is the shorthand notation of the second-order correlation function  $G^{(2)}(r_1=z,r_2=0, t_1=t_2=0;r_1=z,r_2=0,t_1=t_2=0)$. These are the correlation functions for  general thermal light. We would expect $G^{(2)}(z=0)/[G^{(1)}(z=0)]^2=2$, which can be confirmed directly as
\begin{equation}\begin{aligned}
&G^{(2)}(z=0)=\frac{1}{L^2}\sum_{m_1, m_2}2n_{m_1}n_{m_2}=2 [G^{(1)}(z=0)]^2.
\end{aligned}\end{equation}
As $z\rightarrow \infty$, we would expect $G^{(2)}(z\rightarrow\infty)/[G^{(1)}(z\rightarrow\infty)]^2\rightarrow 1$. To prove this, we first note that by the Riemann-Lebesgue Lemma, the Fourier component of a integrable function vanishes as the frequency goes to infinity. We can regard 
$\sum_{m_1, m_2}{n}_{m_1}{n}_{m_2}
e^{i(\kappa_{m_2}-\kappa_{m_1})z}$
as the modulo square of the Fourier transformation of ${n}_{m_1}$, which should go to zero as $z\rightarrow\infty$; thus
\begin{equation}
G^{(2)}(z\rightarrow \infty)=\frac{1}{L^2}\sum_{m_1, m_2}n_{m_1}n_{m_2}=[G^{(1)}(z\rightarrow \infty)]^2.
\end{equation}

If we start from thermal light with a flat spectrum described in the spatial domain as in Eq. \ref{SI_eq:thermal_local},
\begin{equation}
\begin{aligned}\label{flat_G}
&G^{(1)}(z)=N{n}/L,\\
&G^{(2)}(z)=N^2{n}^2/L^2+{n}^2\frac{N}{L}\omega^2(z),
\end{aligned}
\end{equation}
where $\omega(z)=\frac{1}{\sqrt{NL}}\frac{\sin{\pi z/l}}{\sin{\pi z/L}}$. Note if we choose a flat spectrum, i.e. ${n}_m={n}$ in Eq. \ref{general_G}, Eq. \ref{general_G} will be exactly equal to Eq. \ref{flat_G}. One can also easily check that $G^{(2)}(z=0)=2 [G^{(1)}(z=0)]^2$ and $G^{(2)}(z\rightarrow \infty)=[G^{(1)}(z\rightarrow \infty)]^2$, as $\omega(z)$ has the shape of a sinc function as $L\rightarrow\infty$: $\omega(z)\approx \sqrt{N/L} \sin (\pi z/l)/(\pi z /l)$.

For thermal light with a flat spectrum in the weak limit described by Eq. \ref{SI_eq:thermal_weak}, we need to further include the $O(\epsilon^2)$ terms to get the correct $G^{(2)}$. Starting from Eq. \ref{SI_eq:thermal_local}, 
\begin{equation}
\begin{aligned}\label{thermal_local2}
&\rho=\int\left(\prod_s \frac{d^2\gamma_s}{\pi n}\right)\exp(-\sum_s|\gamma_s|^2/n)\left(\otimes_s\ket{\gamma_s}\bra{\gamma_s}\right),
\end{aligned}
\end{equation}
we will keep 
\begin{equation}
\begin{aligned}
&\otimes_s\ket{\gamma_s}=\exp(-\sum_s|\gamma_s|^2/2)\bigg(\ket{0}+\sum_s\gamma_s\ket{1}_s+\frac{1}{2}\sum_{s\neq s'}\gamma_s\gamma_{s'}\ket{1}_s\ket{1}_{s'}+\frac{1}{\sqrt{2}}\sum_s\gamma_s^2\ket{2}_s+o(\gamma_{s}^2)\bigg),
\end{aligned}
\end{equation}
where higher-order terms $o(\gamma_s^2)$ involve at least three photons and contribute nonzero terms to $G^{(2)}$, but the contributions are negligible compared to the $O(\gamma_s^2)$ terms. We can then calculate the correlation functions
\begin{equation}
\begin{aligned}\label{weak_G}
&G^{(1)}(z)=\frac{{n}}{({n}+1)^{N+1}}\frac{N}{L},\\
&G^{(2)}(z)=\frac{{n}^2}{({n}+1)^{N+2}}\left(\frac{N^2}{L^2}+\frac{N}{L}\omega^2(z)\right).
\end{aligned}
\end{equation}
If ${n}\rightarrow0$, this result is consistent with Eq.~\ref{flat_G}. So, all of the correlation functions are consistent with each other.
This is expected because Eqs.~\ref{SI_eq:thermal_local} and \ref{SI_eq:thermal_weak} are the exact decompositions for thermal light with a flat spectrum; their correlation functions will naturally be consistent with those of a thermal state described in the frequency domain. Similarly, we also have $G^{(2)}(z=0)=2 [G^{(1)}(z=0)]^2$, $G^{(2)}(z\rightarrow \infty)=[G^{(1)}(z\rightarrow \infty)]^2$ when $n\rightarrow0$. 

\subsection{Discussion of the trace-improper density matrix in Ref. \cite{chenu2015thermal,chenu2015first}}

We now comment further on the connection between our work and Refs.~\cite{chenu2015thermal, chenu2015first}.
The argument that blackbody radiation cannot be decomposed as a mixture of single pulses is based on comparisons of the first- and second-order correlation functions corresponding to thermal light and a mixture of independent localized pulses \cite{chenu2015thermal,chenu2015first}.  References~\cite{chenu2015thermal,chenu2015first} consider a mixture of single pulses $\rho=\sum_s \sigma_s/N$, where $\sigma_s$ are spatially localized states, such as Gaussian-like pulses. Consider a volume of space $\Omega$ filled with spatially localized pulses with widths $\sigma_s$ independent of $\Omega$. Since blackbody radiation is not localized and fills the whole volume $\Omega$, as we increase $\Omega$, more localized pulses $\sigma_s$ are required to mimic the behavior of blackbody radiation. Consider the first-order correlation function $G^{(1)}(r_1 t_1; r_2 t_2) = \left\langle E^{(-)}(r_1, t_1) E^{(+)}(r_2, t_2) \right\rangle$. If we consider the first-order correlation function $G^{(1)}(0)$, where $t_1=t_2=t$, $r_2-r_1=0$ (for the case of a thermal state, this is the intensity at $\vec{r}=0,t=0$), only a few pulses localized around $\vec{r}=0$ will contribute to $G^{(1)}(0)$ regardless of the total number of localized pulses. This means as $\Omega\rightarrow \infty$, $N\rightarrow \infty$, and hence $G^{(1)}(0)\rightarrow 0$. To fix this vanishing $G^{(1)}(0)$, Ref.~\cite{chenu2015thermal,chenu2015first} introduced a trace-improper and unphysical density matrix. A vanishing $G^{(1)}(0)$ is not a problem for the spectrally flat state of Eq.~\ref{SI_eq:thermal_local}, since it is a direct product instead of the sum of many localized pulses, and so as $\Omega$ is increased, $G^{(1)}(0)$ is not affected. For the spectrally flat state in the weak strength limit of Eq.~\ref{SI_eq:thermal_weak}, assuming $\epsilon\ll 1/N$, terms higher order in $\epsilon$ will not significantly contribute to the first-order correlation function. In this case, $O(\epsilon)$ terms give the correct first-order correlation function. The number of $O(\epsilon)$ terms increases as the volume $\Omega$ increases, similar to the behavior of the trace-improper density matrix in Ref.~\cite{chenu2015thermal,chenu2015first}, except in the limit $\epsilon\ll 1/N$, the trace of the whole density matrix in Eq.~\ref{SI_eq:thermal_weak} is naturally maintained. 

For the second-order correlation function, Refs.~\cite{chenu2015first,chenu2015thermal} found that the decomposition yields a vanishing $G^{(2)}(R)$ as $R \rightarrow \infty$. The correlation function is defined as $G^{(2)}(r_1 t_1, r_2 t_2; r_3 t_3, r_4 t_4) = \left\langle E^{(-)}(r_1, t_1) E^{(-)}(r_2, t_2) E^{(+)}(r_3, t_3) E^{(+)}(r_4, t_4) \right\rangle$, with the specific choice $r_1 = r_4$ and $r_2 = r_3$, $t_1=t_2=t_3=t_4$, $R=r_2-r_1$. An intuitive interpretation is that the decomposition $\rho=\sum_s \sigma_s/N$ allows for only one localized pulse $\sigma_s$ to be present in spatial location $s$. As a localized pulse, $\sigma_s$ cannot provide photons at two well-separated locations simultaneously. This issue is naturally addressed in Eq.~\ref{SI_eq:thermal_local}, which comprises localized pulses distributed over the whole volume $\Omega$. If $R$ is small, there exist correlations since $R$ is within the same localized pulse. If $R\rightarrow\infty$, $g^{(2)}(R)=G^{(2)}(R)/\left(G^{(1)}(R)G^{(1)}(0)\right)=1$ since $R$ spans independent localized pulses in different locations. For the weak-limit expansion in Eq.~\ref{SI_eq:thermal_weak}, the contribution from $O(\epsilon)$ terms vanishes. Therefore, $O(\epsilon^2)$ terms must be included to obtain the correct $G^{(2)}(R)$ as given in Eq.~\ref{weak_G}.

\subsection{Deviation from a flat spectrum} 

\begin{figure}[!hbt]
\begin{center}
\includegraphics[width=.35\columnwidth]{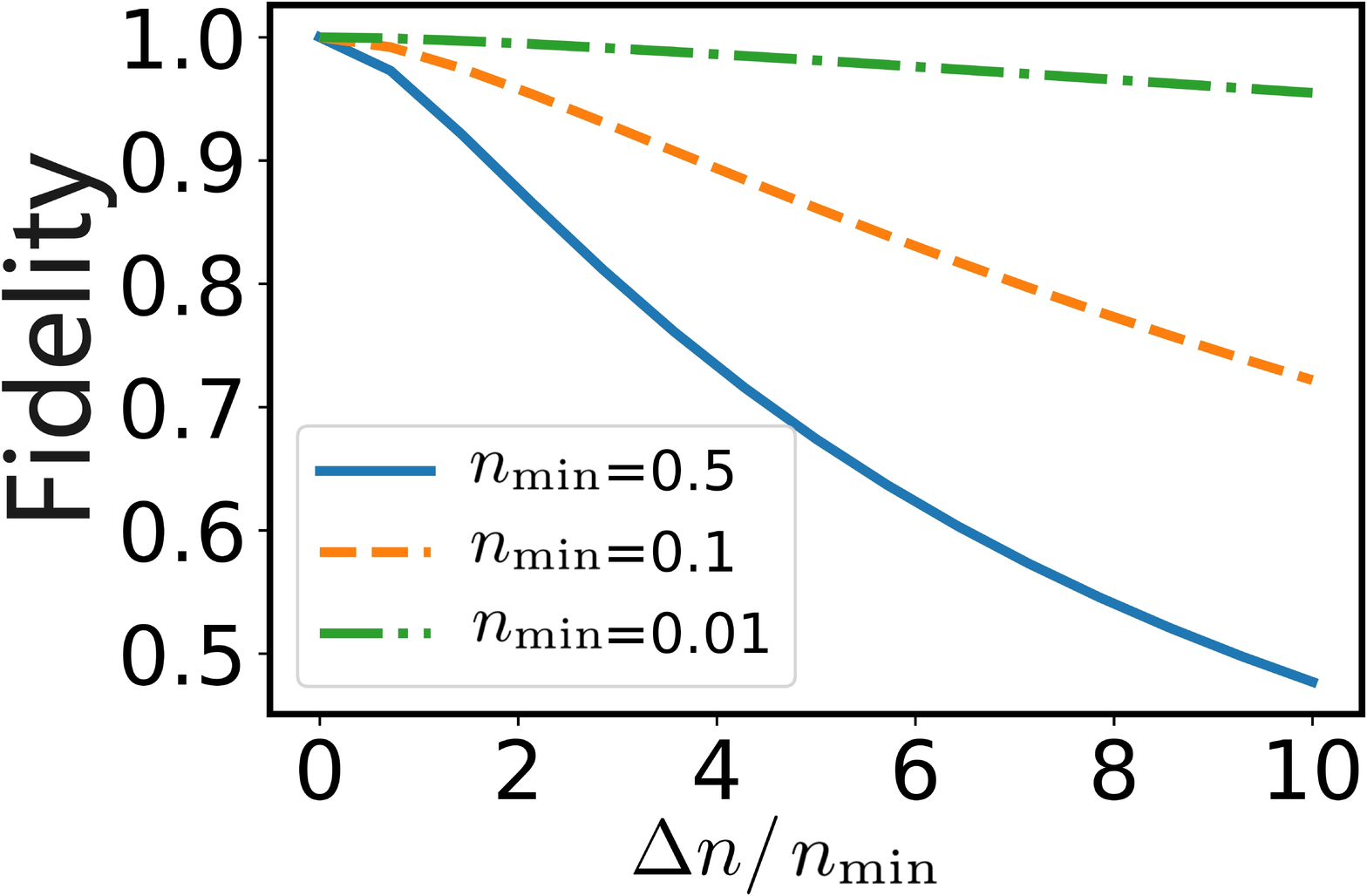}
\caption{The fidelity between a thermal state with a linearly sloped spectrum, where the mean photon number in the frequency modes $n_m$ increases linearly from $n_{\text{min}}$ to $n_{\text{max}}$, and a thermal state with a flat spectrum, as defined in Eq.~\ref{SI_eq:thermal_local}, where the mean photon number for all frequency modes is $n = (n_{\text{min}} + n_{\text{max}})/2$. 
Three cases are plotted: $n_\text{min}=0.01, 0.1, 0.5$, $\Delta n=n_{\text{max}}-n_{\text{min}}$. States with smaller $n_{\text{min}}$ achieve relatively higher fidelity due to greater dominance of vacuum contributions.} 
\label{linear}
\end{center}
\end{figure}

\begin{figure}[!hbt]
\begin{center}
\includegraphics[width=.35\columnwidth]{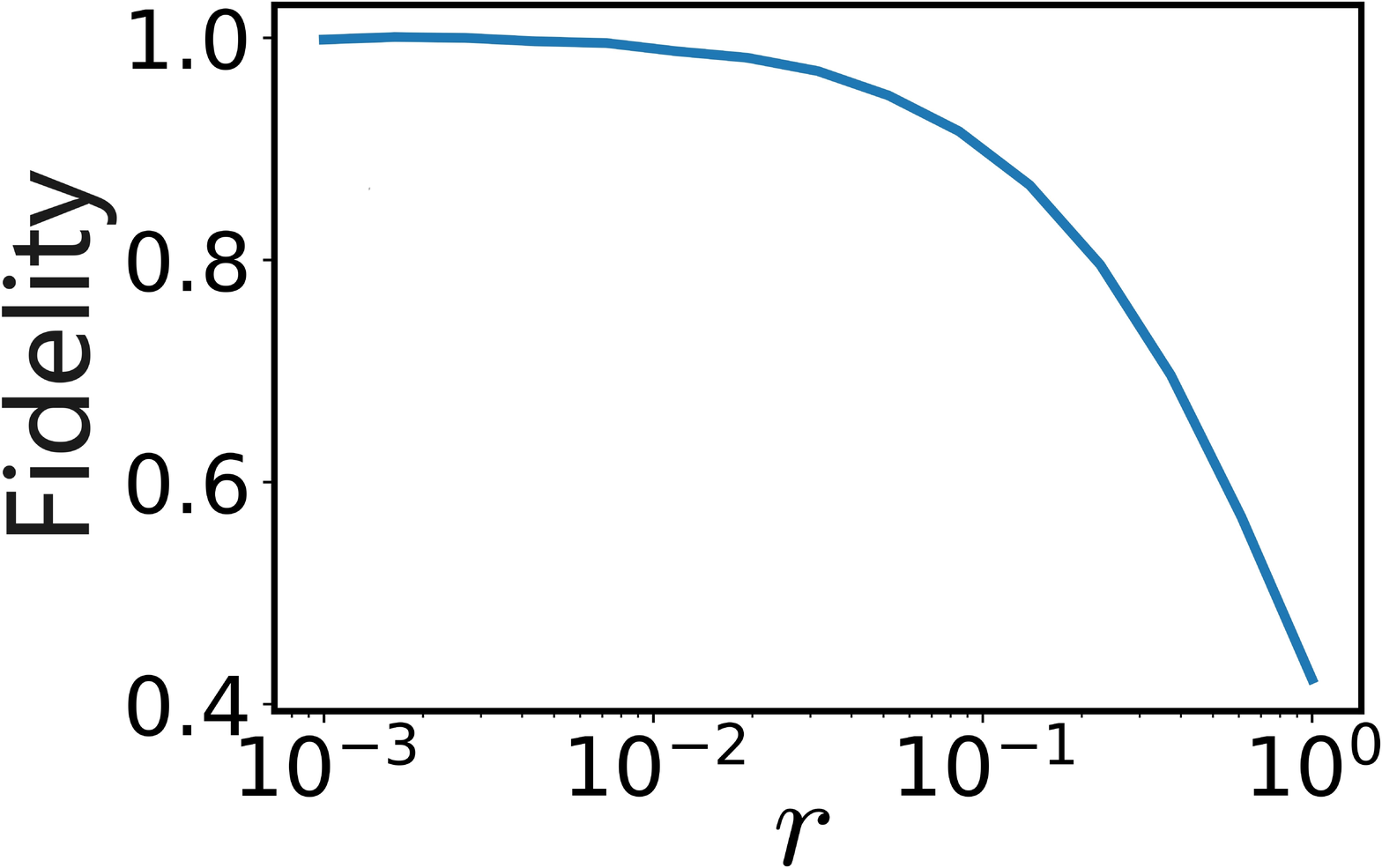}
\caption{The fidelity between a thermal state with a Gaussian spectrum, where the mean photon number in frequency modes $n_m$ follows the distribution $r\exp(-(\omega-\omega_0)^2)/\sqrt{\pi}$ over the range $\omega \in [\omega_0 - 4, \omega_0 + 4]$ with equally spaced modes, and a thermal state with a flat spectrum, as defined in Eq.~\ref{SI_eq:thermal_local}, where the mean photon number for all frequency modes is $n$.} 
\label{gaussian}
\end{center}
\end{figure}

We have focused on a perfectly flat spectrum in the above discussion. In practice, applying a band-pass filter to blackbody light may result in a spectrum that is not perfectly flat. Here we explore two simple cases where the spectrum deviates from flatness. First, we consider a state with a linearly sloped spectrum, where the mean photon number of frequency modes $n_m$ increases from $n_{\text{min}}$ to $n_{\text{max}}$. This spectrum represents a specific instance of the general thermal state described in Eq.~\ref{SI_eq:thermal_state}. We calculate the fidelity between this thermal state and a thermal state with a flat spectrum, as defined in Eq.~\ref{SI_eq:thermal_local}, where the mean photon number for all modes is $n = (n_{\text{min}} + n_{\text{max}})/2$. The result is shown in Fig.~\ref{linear}. We observe that as $\Delta n=n_{\text{max}}-n_{\text{min}}$ increases, the fidelity decreases as expected. The plot indicates a state with a linearly sloped spectrum remains close to the decomposition of thermal light into independent localized pulses provided the deviation $\Delta n$ is not too large. Specifically, the fidelity stays above 90\% provided that $n_{\text{max}} - n_{\text{min}}$ is less than $n_{\text{min}}$ for $n_{\text{min}} = 0.01, 0.1, 0.5$. An interesting observation is that smaller values of $n_{\text{min}},n_{\text{max}}$ result in better fidelity for a given $\Delta n/n_{\text{min}}$. We think this is due to the fact that as $n_m\rightarrow 0$, the fidelity calculation becomes dominated by vacuum contributions.  Applying appropriate compensation filters can ensure a flat spectrum; this is especially important for stronger sources, for which the fidelity is more sensitive to the slope.  The second case we consider is a state with a Gaussian spectrum, where the mean photon number in each frequency mode, $n_m$, follows a Gaussian distribution given by $r \exp[-(\omega - \omega_0)^2]/\sqrt{\pi}$. Here, $r$ is a positive scalar that controls the overall intensity of the source across all frequencies. This case could correspond to the shape of the filtered spectrum being dominated by the width of the filter's edge transitions, which results in an approximately Gaussian shape. Alternatively, this case could represent a natural source with a Gaussian spectrum. For comparison with a thermal state with a flat spectrum as in Eq.~\ref{SI_eq:thermal_local}, we numerically optimize the choice of the mean photon number $n$ for all frequency modes in order to obtain the best fidelity for a given $r$. The results are shown in Fig.~\ref{gaussian}. For $r$ smaller than $10^{-1}$, corresponding to a low mean photon number, the fidelity remains above 90\%. These calculations reveal that the decomposition of thermal light into independent localized pulses exhibits a certain degree of robustness and remains relatively stable under slight deviations from a flat spectrum.

\section{Astronomical interferometry of thermal light with a general spectrum}\label{SI:astronomical_interferometry}

\subsection{Decomposition of the states received by an astronomical interferometer into mixtures of localized pulses}

We now provide further details on imaging thermal sources using an astronomical interferometer in the case of a general spectrum.
Let's start from the P representation of a general incoherent source
\begin{equation}
\begin{aligned}
&\rho=\int d^{2MW}\vec{\alpha}\Phi(\vec{\alpha})\ket{\vec{\alpha}}\bra{\vec{\alpha}},\quad\ket{\vec{\alpha}}=\otimes_{i=1}^W\otimes_{m=1}^M\exp(\alpha_{im}a_{i,k_m}^\dagger-\alpha_{im}^*a_{i,k_m})\ket{0}\\
&\Phi(\vec{\alpha})=\frac{\exp(-\vec{\alpha}^\dagger\Gamma^{-1}\vec{\alpha})}{\det(\pi\Gamma)},\quad \Gamma_{i_1m_1,i_2m_2}=\delta_{i_1i_2}\delta_{m_1m_2}n_{i_1m_1},
\end{aligned}
\end{equation}
where $i = 1, 2, \ldots, W$ labels the $i$th point on the source, and $m = 1, 2, \ldots, M$ labels different frequencies $k_m$. We take the limits $W, M \to \infty$ to model a continuous spatial intensity distribution and a continuous spectrum.
The quantity $n_{im}$ denotes the mean photon number emitted at frequency $k_m$ from the $i$th point on the source. For simplicity, we consider a one-dimensional thermal source by neglecting variation along the $y$-axis. The spatial coordinate of the $i$th source point is taken to be $(x_i, d)$, where $x_i$ is the position along the $x$-axis and $d$ is the fixed distance between the source and the telescope array along the $z$-axis.

We consider the case of two telescopes. Since the astronomical source is in the far-field limit relative to the telescopes on Earth, we assume that the light near the telescopes propagates only along the $z$-direction. Let the two lenses be located at positions $(u_0, 0)$ and $(u_1, 0)$. Under this setup, the annihilation operators evolve as follows:
\begin{equation}\begin{aligned}
a_{i,k_m}\rightarrow &\int dz \sqrt{\eta}e^{i\phi(k_m,u_0,z,x_i)}\hat{b}_{0z}+\int dz \sqrt{\eta}e^{i\phi(k_m,u_1,z,x_i)}\hat{b}_{1z}+\hat{v}_{i,k_m},    
\end{aligned}
\end{equation}
where $\hat{v}_{i,k_m}$ represents environmental modes, and $\eta$ quantifies the transmission loss from the source to the spatial modes near the two telescopes. The operators $\hat{b}_{0z}$ and $\hat{b}_{1z}$ correspond to the spatial modes just above the first and second telescopes, located at positions $(u_0, z)$ and $(u_1, z)$, respectively. The phase accumulated along the path from a point on the source at $(x,d)$ to the position $(u,z)$ above the telescopes
\begin{equation}\begin{aligned}
&\phi(k,u,z,x)=k\sqrt{(x-u)^2+(d-z)^2}=kd\left[1-\frac{z}{d}+\frac{x^2+u^2}{2d^2}-\frac{xu}{d^2}+O(\frac{x^3+u^3+z^3}{d^3})\right],  
\end{aligned}
\end{equation}
where we assume $d\gg x,u,z$. With the above analysis, let us find the states near the two telescopes.
\begin{equation}\begin{aligned}
&\ket{\vec{\alpha}}\rightarrow\ket{\vec{\beta}},\quad\vec{\beta}=O\vec{\alpha}\\
&\ket{\vec{\beta}}=\exp(\sum_{n=1}^N\sum_{p=0}^1[\beta_{pn}\hat{b}_{p,z_n}^\dagger-\beta_{pn}^*\hat{b}_{p,z_n}])   \\
&\beta_{pn}=\sum_{i=1}^W\sum_{j=1}^M O_{pn,ij}\alpha_{ij}\\
&O_{pn,ij}=\sqrt{\eta}e^{-i\phi(k_j,u_p,z_n,x_i)}
\end{aligned}
\end{equation}
where $p=0,1$, $n=1,2,\cdots,N$. We can then find the form of states near the telescopes: 
\begin{equation}
\begin{aligned}
&\rho=\int d^{4N}\Phi(\vec{\beta})\ket{\vec{\beta}}\bra{\vec{\beta}},\quad\Phi(\vec{\beta})=\frac{\exp(-\vec{\beta}^\dagger\Lambda^{-1}\vec{\beta})}{\det(\pi\Lambda)},\\
&\Lambda_{lt,pq}=\sum_{i=1}^W \sum_{m=1}^M \eta n_{im}\exp(i\phi(u_p,z_q,x_i,k_m)-i\phi(u_l,z_t,x_i,k_m))\\
&=\sum_{i=1}^W \sum_{m=1}^M \eta n_{im}\exp \bigg(ik_m(z_t-z_q)+ik_m\frac{u_p^2-u_l^2}{2d}+ik_m\frac{x_i(u_l-u_p)}{d}+O(\frac{x^3+u^3+z^3}{d^3})\bigg).
\end{aligned}
\end{equation}

We now show how, in the weak-light limit, the state $\rho$ can be decomposed as a mixture of localized pulses. Define $\vec{p}, \vec{q} \in \{0,1\}^{\otimes Q}$ to label which telescope each photon is associated with. Let $\vec{s} = [s_1, s_2, \ldots, s_Q]$, where each $s_i \in \{1, 2, \ldots, N\}$ denotes the temporal mode in which the $i$th photon is detected. 
\begin{equation}
\ket{\vec{s},\vec{p}}=A(\vec{s},\vec{p})\sum_{n_1,n_2,\cdots,n_Q=1}^N\left(\prod_{i=1}^Q\omega_{s_i}(z_{n_i})\hat{b}_{p_i z_{n_i}}^\dagger\right)\ket{0},
\end{equation}
where the prefactor is given by $A(\vec{s},\vec{p}) = 1 / \sqrt{\prod_{j=1}^N (K_{j0}!)(K_{j1}!)}$, with $K_{j0}$ denoting the number of occurrences of $s_i = j$, $p_i=0$, $K_{j1}$ denoting the number of occurrences of $s_i = j$, $p_i=1$, i.e., the number of photons in the $j$th temporal mode at each telescope. Similarly, we define the state $\ket{\vec{s}', \vec{q}}$ in the same way, with $\vec{s}'$ labeling a possibly different set of temporal modes and $\vec{q}$ indicating the corresponding telescope indices. 
\begin{equation}\begin{aligned}
&\bra{\vec{s'},\vec{p}}\rho\ket{\vec{s},\vec{q}}=\int d^{4N}\vec{\beta}\frac{\exp(-\vec{\beta}^\dagger\Lambda^{-1}\vec{\beta})}{\det(\pi\Lambda)}\exp(-\vec{\beta}^\dagger\vec{\beta})\\
&\quad\quad\quad\times A(\vec{s'},\vec{p})\left(\sum_{n_1,n_2,\cdots,n_Q=1}^N\prod_{i=1}^Q \omega_{s_i'}^*(z_{n_i})\beta_{p_i z_{n_i}}\right)A(\vec{s},\vec{q})\left(\sum_{m_1,m_2,\cdots,m_Q=1}^N\prod_{i=1}^Q \omega_{s_i}(z_{m_i})\beta_{q_i z_{m_i}}^*\right)\\
&\approx A(\vec{s},\vec{p})A(\vec{s'},\vec{q})\sum_\mathbb{P}\prod_{i=1}^Q\left(\sum_{n_i=1}^N\sum_{m_{\mathbb{P}_i}=1}^N\omega_{s_i'}^*(z_{n_i}) \Lambda_{p_i n_i,q_{\mathbb{P}_i}m_{\mathbb{P}_i}}\omega_{s_{\mathbb{P}_i}}(z_{m_{\mathbb{P}_i}})\right)\\
&=A(\vec{s},\vec{p})A(\vec{s'},\vec{q})\sum_\mathbb{P}\left(\prod_{i=1}^Q\vec{\omega}_{s_i' p_i}^\dagger \Lambda\vec{\omega}_{s_{\mathbb{P}_i}q_{\mathbb{P}_i}}\right),\\
\end{aligned}\end{equation}
where the sum is taken over all permutations $\mathbb{P}$, and we work in the weak-source limit $\lambda_{\text{max}}(\Lambda) \rightarrow 0$. The vector $\vec{\omega}_{s_i p_i}$ denotes the $i$th measurement mode at the first or second telescope introduced for convenience of derivation, specified by $p_i \in \{0,1\}$ and the $s_i$th temporal mode. Under these assumptions, the state admits the following decomposition:
\begin{equation}
\begin{aligned}
&\rho=\sum_{Q=0}^\infty\sum_{\vec{s},\vec{s'}}\sum_{\vec{p},\vec{q}} A(\vec{s},\vec{q})A(\vec{s'},\vec{p})\sum_P\left(\prod_{i=1}^Q\vec{\omega}_{s'_i p_i}^\dagger \Lambda \vec{\omega}_{s_{P_i}q_{P_i}}\right)\ket{\vec{s'},\vec{p}}\bra{\vec{s},\vec{q}}\\
&=\ket{0}\bra{0}+\sum_{s,s'=1}^N\sum_{p,q\in\{0,1\}}\vec{\omega}_{s' p}^\dagger \Lambda \vec{\omega}_{sq}\ket{s',p}\bra{s,q}\\
&+\sum_{\vec{s},\vec{s'}}\sum_{\vec{p},\vec{q}\in\{0,1\}^{\otimes 2}}A(\vec{s},\vec{q})A(\vec{s'},\vec{p})\sum_\mathbb{P}\prod_{i=1}^2\vec{\omega}_{s'_i p_i}^\dagger \Lambda\vec{\omega}_{s_{\mathbb{P}_i}q_{\mathbb{P}_i}}\ket{\vec{s'},\vec{p}}\bra{\vec{s},\vec{q}}+\cdots
\end{aligned}
\end{equation}
Note that $\vec{s}$ is a $Q$-dimensional vector. We observe that the state decomposes into a mixture of localized pulses that contain photons. In general, there are correlations between different temporal modes.

\subsection{Correction factor for non-flat spectrum}

We now investigate the effect of a non-flat spectrum on the actual successive measurements. Although in the main text we first introduce Proposition 1 followed by Proposition 2—for logical clarity—we present the proof of Proposition 2 first, as it follows a slightly more natural mathematical flow.

In practical astronomical interferometry at optical wavelengths, the weak-source limit typically applies. Suppose we perform $N$ successive measurements and detect a total of $Q$ photons across the temporal modes $\vec{s} = [s_1, s_2, \ldots, s_Q]$, where $s_i=1,2,\cdots,N$ indicates that a photon is detected in the $s_i$th temporal mode. The measurement corresponding to a single-photon detection in the $s_i$th temporal mode projects the state onto 
\begin{equation}\begin{aligned}
&\{\ket{00}\bra{00}_{s_i},\ket{\pm e^{i\theta}}\bra{\pm e^{i\theta}}\},\quad\ket{\pm e^{i\theta}}=(\ket{01}_{s_i}\pm e^{i\theta}\ket{10}_{s_i})/\sqrt{2}\\   
&\ket{01}_{s_i}=\sum_{n=1}^N\omega_{s_i}(z_n)b_{0z_n}^\dagger\ket{0},\quad\ket{10}_{s_i}=\sum_{n=1}^N\omega_{s_i}(z_n)b_{1z_n}^\dagger\ket{0},
\end{aligned}
\end{equation}
where ${\omega_i(z)}_{i=1,2,\ldots,\infty}$ is a set of orthonormal, spatially localized basis functions. We define $\vec{\delta} = \{e^{i\theta}, -e^{i\theta}\}^{\otimes Q}$ to label the outcomes of $Q$ measurements, each of which  involves exactly one detected photon. The measurement process projects the state onto
\begin{equation}
\ket{\vec{s},\vec{\delta}}=\sum_{\vec{p}\in\{0,1\}^{\otimes Q}}\left(\prod_{i=1}^Q\delta_i^{p_i}\right)\ket{\vec{s},\vec{p}}/2^{Q/2},
\end{equation}
which has probability distribution
\begin{equation}\begin{aligned}\label{eq:P_full}
&P(\vec{s},\vec{\delta})=\bra{\vec{s},\vec{\delta}}\rho\ket{\vec{s},\vec{\delta}}=\frac{1}{2^Q}\sum_{\vec{p},\vec{q}\in\{0,1\}^{\otimes Q}}\left[\left(\prod_{i=1}^Q\delta_i^{p_i+q_i}\right)\sum_\mathbb{P}\left(\prod_{i=1}^Q \vec{\omega}_{s_i p_i}^\dagger\Lambda\vec{\omega_{s_{\mathbb{P}_i}q_{\mathbb{P}_i}}}\right)\right].
\end{aligned}
\end{equation} 
To prove Proposition 2, we can perform the following simplification  by applying the simplification based on the assumption $n_{im} = I_i f_m$ and the identity $\sum_{n=1}^N \omega_s(z_n)\exp(-ik_m z_n) = \tilde{\omega}_s(k_m)$
\begin{equation}\begin{aligned}
&\vec{\omega}_{s_1,p}^\dagger\Lambda\vec{\omega}_{s_2,q}=\sum_{n_1,n_2=1}^N\omega_{s_1}^*(z_{n_1})\Lambda_{pn_1,qn_2}\omega_{s_2}(z_{n_2})\\
&=\sum_{n_1,n_2=1}^N\sum_{i=1}^W\sum_{m=1}^M\omega_{s_1}^*(z_{n_1})\omega_{s_2}(z_{n_2})\eta n_{im}\exp\left(ik_m(z_{n_1}-z_{n_2})+ik_m\frac{u_q^2-u_p^2}{2d}+ik_m\frac{x_i(u_p-u_q)}{d}\right)\\
&=\sum_{m=1}^Mf_m\tilde{\omega}_{s_1}^*(k_m)\tilde{\omega}_{s_2}(k_m)\sum_{i=1}^W\eta I_i\exp\left(ik_m\frac{u_q^2-u_p^2}{2d}+ik_m\frac{x_i(u_p-u_q)}{d}\right)
\end{aligned}
\end{equation}
We define the overlap between the spectra of two different temporal modes, $\tilde{\omega}_{s_1}(k_m)$ and $\tilde{\omega}_{s_2}(k_m)$, weighted by the source spectrum $f_m$
\begin{equation}\begin{aligned}
\Omega_{s_1s_2}=\sum_{m=1}^Mf_m\tilde{\omega}^*_{s_1}(k_m)\tilde{\omega}_{s_2}(k_m).
\end{aligned}
\end{equation}
The matrix $\Omega_{s_1 s_2}$ becomes diagonal if the temporal modes are orthogonal with respect to the inner product weighted by the spectral distribution $f_m$. In the case of a flat spectrum, since the functions $\tilde{\omega}(k)$ form an orthonormal set, we expect $\Omega_{s_1 s_2}$ to vanish for $s_1 \neq s_2$. Therefore, $\Omega_{s_1 s_2}$ can be understood as a measure of temporal correlations between modes $s_1$ and $s_2$ in the presence of a non-flat spectrum. We define
\begin{equation}\begin{aligned}
\epsilon =\sum_{i=1}^W\eta I_i,\quad\sum_{i=1}^W\eta I_i\exp\left(ik_m\frac{u_q^2-u_p^2}{2d}+ik_m\frac{x_i(u_p-u_q)}{d}\right)=\epsilon g_m(p,q),\quad g_m(p,q)=\left\{ 
  \begin{array}{ll}
     g_m & \text{if } p=0,q=1 \\
     g_m^*  & \text{if } p=1,q=0\\
    1 & \text{if } p=q=0,1\\
  \end{array}
\right.,
\end{aligned}
\end{equation}
where $g_m$ denotes the coherence function at frequency $k_m$, and $\epsilon$ quantifies the source strength. For simplicity, we assume $g_m = g$ is constant for all $m$. By definition, $g_m$ corresponds to the spatial Fourier component of the source’s intensity distribution at spatial frequency $k_m(u_p - u_q)/d$. Different $k_m$ correspond to different spatial frequencies.
In our analysis, we assume $k_m \in [k_0 - \Delta k/2,, k_0 + \Delta k/2]$ with $\Delta k \ll k_0$. Thus, if the Fourier transform of the source intensity is sufficiently smooth, this constant-$g_m$ approximation is well justified. This assumption is adopted primarily to facilitate the following analytical treatment.
If one wishes to incorporate spectral variation in $g_m$, we can perform a Taylor expansion around $k_0$: $g_m = g + a(k_m - k_0) + b(k_m - k_0)^2 + \cdots$, where the coefficients $a$, $b$, etc., are independent of $k_m$. Correspondingly, we can define generalized overlap coefficients analogous to $\Omega_{s_1 s_2}$, $\sum_{m=1}^M f_m(k_m-k_0)^l \tilde{\omega}^*_{s_1}(k_m)\tilde{\omega}_{s_2}(k_m)$, which can be directly evaluated, as both the spectral profile $f_m$ and the detection-mode shapes $\tilde{\omega}_s(k_m)$ are known. These quantities can be obtained through calibration measurements, given the assumption $n_{im} = I_i f_m$, without requiring knowledge of the spatial intensity distribution $I_i$.

With the above assumption that $g_m=g$, we can find the following form  relevant to the terms in the probability distribution associated with the temporal modes $s_1$ and $s_2$
\begin{equation}\begin{aligned}\label{SI_eq:g_pq}
&\vec{\omega}_{s_1,p}^\dagger\Lambda\vec{\omega}_{s_2,q}=\epsilon \Omega_{s_1s_2}g(p,q),\quad g(p,q)=\left\{ 
  \begin{array}{ll}
    g & \text{if } p=0,q=1 \\
     g^*  & \text{if } p=1,q=0\\
    1& \text{if } p=q=0,1\\
  \end{array}
\right..
\end{aligned}
\end{equation}
We can then write the probability distribution as
\begin{equation}
\begin{aligned}
&P(\vec{s},\vec{\delta})=\bra{\vec{s},\vec{\delta}}\rho\ket{\vec{s},\vec{\delta}}=\frac{1}{2^Q}\sum_{\vec{p},\vec{q}\in\{0,1\}^{\otimes Q}}\left[\left(\prod_{i=1}^Q\delta_i^{p_i+q_i}\right)\sum_P\left(\prod_{i=1}^Q \vec{\omega}_{s_i p_i}^\dagger\Lambda\vec{\omega_{s_{P_i}q_{P_i}}}\right)\right]\\
&=\frac{\epsilon^Q}{2^Q}\sum_{\vec{p},\vec{q}\in\{0,1\}^{\otimes Q}}\sum_{\mathbb{P}}\prod_{i=1}^Q(\delta_i^{p_i+q_i}g(p_i,q_{\mathbb{P}_i})\Omega_{s_i,s_{\mathbb{P}_i}})\\
&=\frac{\epsilon^Q}{2^Q}\sum_{\mathbb{P}}\prod_{i=1}^Q[(\delta_ig^*+\delta_{\mathbb{P}_i}g+1+\delta_i\delta_{\mathbb{P}_i})\Omega_{s_i,s_{\mathbb{P}_i}}].
\end{aligned}
\end{equation}

Let us consider two examples to examine how the probability distribution is affected in the case of a non-flat spectrum.
When $Q=2$, $\vec{s}=[1,2]$, $\vec{\delta}=[1,-1]$
\begin{equation}
\begin{aligned}\label{SI_eq:two_photon_P}
&P(\vec{s},\vec{\delta})=\frac{\epsilon^2}{4}(g+g^*+2)(-g-g^*+2)\Omega_{11}\Omega_{22}+\frac{\epsilon^2}{4}(-g+g^*)(g-g^*)\Omega_{12}\Omega_{21},
\end{aligned}
\end{equation}
where the second term will influence the estimation outcome if the product $\Omega_{12} \Omega_{21}$ is sufficiently large.

When $Q=3$, $\vec{s}=[1,2,3]$, $\vec{\delta}=[1,1,-1]$
\begin{equation}
\begin{aligned}
P(\vec{s},\vec{\delta})=\frac{\epsilon^3}{8}&[(g+g^*+2)^2(-g-g^*+2)(\Omega_{11}\Omega_{22}\Omega_{33}+\Omega_{12}\Omega_{21}\Omega_{33})\\
&-(g+g^*+2)(g-g^*)^2(\Omega_{13}\Omega_{22}\Omega_{31}+\Omega_{11}\Omega_{23}\Omega_{32}+\Omega_{12}\Omega_{23}\Omega_{31}+\Omega_{13}\Omega_{21}\Omega_{32})],
\end{aligned}
\end{equation}
where additional terms arise as a result of correlations between different temporal modes.

To estimate $|g|$ or $\theta = \arg g$, one possible approach is to use the maximum likelihood estimator (MLE). To compute the estimate $\hat{|g|}$, we can simply evaluate
\begin{equation}
\frac{\partial{P(\vec{s},\vec{\delta})}}{\partial |g|}\bigg|_{|g|=\hat{|g|}}=0, \quad \frac{\partial{P(\vec{s},\vec{\delta})}}{\partial \theta}\bigg|_{\theta=\hat{\theta}}=0,
\end{equation}
for a given outcome $\vec{s}, \vec{\delta}$. Note that to estimate both parameters $\theta$ and $|g|$, it is necessary to vary the measurement settings by introducing controlled phase delays, effectively modifying each $\delta_i = \pm e^{i\theta}$. Similarly, if we wish to include higher-order terms in the expansion $g_m = g + a(k_m - k_0) + b(k_m - k_0)^2 + \cdots$, additional phase delays must be applied to extract these extra unknown coefficients.

\subsection{Criteria for neglecting the correlation between temporal modes}

We now prove Proposition 1, which quantifies the difference between the actual probability distribution $P(\vec{s}, \vec{\delta})$ and the idealized distribution $P_0(\vec{s}, \vec{\delta})$ obtained under the assumption of no temporal correlations.
Note that if different temporal modes are uncorrelated, we expect $\vec{\omega}_{i p_i}^\dagger \Lambda \vec{\omega}_{j q_j} = 0$ for $i \neq j$, which in turn implies that
\begin{equation}\begin{aligned}
&P_0(\vec{s},\vec{\delta})=\bra{\vec{s},\vec{\delta}}\rho\ket{\vec{s},\vec{\delta}}=\frac{1}{2^Q}\sum_{\vec{p},\vec{q}\in\{0,1\}^{\otimes Q}}\left[\left(\prod_{i=1}^Q\delta_i^{p_i+q_i}\right)\left(\prod_{i=1}^Q \vec{\omega}_{s_i p_i}^\dagger\Lambda\vec{\omega_{s_iq_{i}}}\right)\right]\\
&=\prod_{i=1}^Q\left(\frac{1}{2}\sum_{p,q\in\{0,1\}}\delta_i^{p+q}\vec{\omega}_{s_ip}^\dagger\Lambda\vec{\omega}_{s_iq}\right).
\end{aligned}
\end{equation} 
We now quantify the deviation  of the actual probability distribution $P(\vec{s}, \vec{\delta})$, for a non-flat spectrum, from the reference distribution $P_0(\vec{s}, \vec{\delta})$, which neglects all intermode correlations:
\begin{equation}
\begin{aligned}
|P(\vec{s},\vec{\delta})-P_0(\vec{s},\vec{\delta})|=\left|\sum_{\mathbb{P}\neq I}\frac{1}{2^Q}\sum_{\vec{p},\vec{q}\in\{0,1\}^{\otimes Q}}\left(\prod_{i=1}^Q\delta_i^{p_i+q_i}\vec{\omega}_{s_ip_i}^\dagger\Lambda\vec{\omega}_{s_{P_i}q_{\mathbb{P}_i}}\right)\right|\leq \sum_{\mathbb{P}\neq I}\frac{1}{2^Q}\sum_{\vec{p},\vec{q}\in\{0,1\}^{\otimes Q}}\left|\prod_{i=1}^Q \vec{\omega}_{s_i p_i}^\dagger\Lambda\vec{\omega}_{s_{P_i}q_{\mathbb{P}_i}}\right|.
\end{aligned}
\end{equation}
We define the following function to quantify the correlation between the two temporal modes $s_1$ and $s_2$
\begin{equation}
\begin{aligned}
\xi(x_{s_1}-x_{s_2})=\frac{|\vec{\omega}_{s_1 a}^\dagger\Lambda\vec{\omega}_{s_2b}|}{|\vec{\omega}_{s a}^\dagger\Lambda\vec{\omega}_{sb}|}=\frac{\Omega_{s_1s_2}}{\Omega_{ss}}=\frac{\int_{k_0-\Delta k/2}^{k_0+\Delta k/2}dk f(k)\tilde{\omega}_{s_1}^*(k)\tilde{\omega}_{s_2}(k)}{\int_{k_0-\Delta k/2}^{k_0+\Delta k/2}dk f(k)\tilde{\omega}_{s}^*(k)\tilde{\omega}_{s}(k)}.
\end{aligned}
\end{equation}
With the assumption that $\omega_{s}(z)=\omega(z-z_s)$,  we can easily find $\tilde{\omega}_s(k)=e^{-ikx_s}\tilde{\omega}(k)$. For convenience in bounding the behavior of $\xi(x_{s_1} - x_{s_2})$, we further define  a normalized distribution function 
\begin{equation}
p(k)=\frac{ f(k)|\tilde{\omega}(k)|^2}{\int_{k_0-\Delta k/2}^{k_0+\Delta k/2}dk f(k)|\tilde{\omega}(k)|^2},
\end{equation}
which then gives
\begin{equation}
\begin{aligned}\label{SI_eq:error_fun}
&\xi(x_{s_1}-x_{s_2})=\frac{\Omega_{s_1s_2}}{\Omega_{ss}}=\int_{k_0-\Delta k/2}^{k_0+\Delta k/2}dk p(k)\exp(ik\Delta x)\\
&=\left(\frac{p(k)}{i\Delta x}\exp(ik\Delta x)\right)\bigg|_{k_0-\Delta k/2}^{k_0+\Delta k/2}-\frac{1}{i\Delta x}\int_{k_0-\Delta k/2}^{k_0+\Delta k/2}dk p'(k)\exp(ik\Delta x)\\
&\leq \frac{2\|p(k)\|_{\infty}}{|\Delta x|}+\frac{\Delta k\|p'(k)\|_{\infty}}{|\Delta x|},
\end{aligned}
\end{equation}
where $\Delta x = x_{s_1} - x_{s_2}$. Note that since $\int dk\, p(k) = 1$, the function $p(k)$ has units of inverse length and is typically of order $1/\Delta k$. In practice, $p(k)$ can be directly estimated with some calibration overhead. When $\Delta x$ is sufficiently large, the above upper bound can be used to show that the correlation between two temporal modes $\omega_{s_1}(z)$ and $\omega_{s_2}(z)$ becomes negligible.


Using the above definition, we can bound the deviation of the probability distribution in terms of the functions $\xi(x_{s_1} - x_{s_2})$, which characterize the correlations between different temporal modes.
\begin{equation}
\begin{aligned}\label{eq:P_difference}
|P(\vec{s},\vec{\delta})-P_0(\vec{s},\vec{\delta})|\leq \sum_{\mathbb{P}\neq I}\frac{1}{2^Q}\sum_{\vec{p},\vec{q}\in\{0,1\}^{\otimes Q}}\left|\prod_{i=1}^Q\xi(x_{s_i}-x_{s_{\mathbb{P}_i}} )\vec{\omega}_{s_i p_i}^\dagger\Lambda\vec{\omega}_{s_{i}q_{\mathbb{P}_i}}\right|\leq(2\epsilon)^Q\sum_{\mathbb{P}\neq I}\prod_{i=1}^Q\xi(x_{s_i}-x_{s_{\mathbb{P}_i}} )
\end{aligned}
\end{equation}
Obviously, $P_0(\vec{s}, \vec{\delta})$ is also of order $\epsilon^Q$. Note that the sum excludes the identity permutation, so as long as all photons are detected with sufficiently large time intervals between them, the difference $|P(\vec{s}, \vec{\delta}) - P_0(\vec{s}, \vec{\delta})|$ will be negligible. In practice, $\vec{s}$ and $\vec{\delta}$ represent the detection outcome from measuring $N$ localized temporal modes and registering $Q$ photons. Since the detection spectrum can be obtained from calibration measurements  without knowing the intensity distribution of the source, we can determine the spectral profile $f(x)$, allowing Eq.\ref{eq:P_difference} to be used to assess whether further correction is necessary. If a significant fraction of photons are detected in closely spaced temporal modes, data postprocessing must account for the resulting correlations, as described in Eq.\ref{eq:P_full} and demonstrated in the Sec. \ref{SI:Gaussian_example} of the Supplemental Material with a more detailed example.

The expected value of $Q \approx 2\epsilon N$ for detection over $N$ temporal modes. This can be intuitively observed because
\begin{equation}
\begin{aligned}
P(\vec{s})=\sum_{\vec{\delta}\in\{1,-1\}^{\otimes Q}}P(\vec{s},\vec{\delta})=\sum_{\vec{p}\in\{0,1\}^{\otimes Q}}\sum_\mathbb{P}\prod_{i=1}^Q \vec{\omega}_{s_i p_i}^\dagger\Lambda\vec{\omega}_{s_{\mathbb{P}_i}p_{\mathbb{P}_i}},
\end{aligned}
\end{equation}
where we choose $\vec{\delta}\in\{1,-1\}^{\otimes Q}$ as an example.
As an approximation, we assume the temporal modes are independent,
\begin{equation}
\begin{aligned}
P(\vec{s})=\sum_{\vec{p}\in\{0,1\}^{\otimes Q}}\prod_{i=1}^Q \vec{\omega}_{s_i p_i}^\dagger\Lambda\vec{\omega}_{s_{i}p_{i}}=(2\epsilon)^Q.
\end{aligned}
\end{equation}
The probability of getting $Q$ photons is
\begin{equation}
\begin{aligned}
P(Q)=C_N^Q(2\epsilon)^Q.
\end{aligned}
\end{equation}
By calculating the ratio $P(Q+1)/P(Q) = (N - Q)\epsilon / (Q + 1)$, we see that the probability is maximized at $Q \approx 2N\epsilon$. This is intuitively reasonable, as $\epsilon$ roughly represents the mean photon number detected at each temporal mode for one of the telescopes. When measuring $N$ temporal modes with two telescopes, we are likely to detect approximately $Q \approx 2N\epsilon$ photons.
Thus, for a weak source, we can expect $Q \ll N$, and in most cases the difference $|P(\vec{s}, \vec{\delta}) - P_0(\vec{s}, \vec{\delta})|$, as quantified in Eq.~\ref{eq:P_difference}, is negligible.


\subsection{Examples with Gaussian spectrum and pulse shape}\label{SI:Gaussian_example}

We now present an explicit calculation of the criterion established in our work, along with the corresponding corrections, using a concrete choice of spectrum and pulse shape. Consider the simplest case of detecting two photons at positions $x_{s_1}$ and $x_{s_2}$, respectively, with $\vec{\delta} = [1, -1]$. We choose the spectrum as $f_m = \frac{a}{M} \exp\left( -\frac{(k_m - k_0)^2}{\sigma^2} \right)$, and the pulse shape as $\tilde{\omega}(k_m) = \sqrt{b} \exp\left( -\frac{(k_m - k_0)^2}{2\beta^2} \right)$, which is nonvanishing only within the frequency range $k \in [k_0 - \Delta k/2,, k_0 + \Delta k/2]$. This pulse shape can be approximately realized by applying a Gaussian-shaped bandpass filter in the frequency domain and detecting within a time window of duration $1/\beta$. Note that, due to temporal overlap between Gaussian pulses, this only serves as an approximation to the ideal Gaussian mode considered here, and such overlap effects will be ignored in this illustrative example. This pulse shape can be exactly

Due to this cutoff in frequency, we must normalize the function accordingly. Note that converting the discrete sum to an integral introduces additional prefactors. From the definition of $f_m$, we have
\begin{equation}
1=\sum_{m=1}^Mf_m=\sum_{m=1}^M\frac{a}{M} \exp(-(k_m-k_0)^2/\sigma^2)=\int_{k_0-\Delta k/2}^{k_0+\Delta k/2}\frac{dk}{\Delta k/M} \frac{a}{M} \exp(-(k-k_0)^2/\sigma^2)=\int_{k_0-\Delta k/2}^{k_0+\Delta k/2}dk f(k)
\end{equation}
Thus, we have $a=\Delta k/(\sqrt{\pi}\sigma\,\text{erf}(\Delta k/2\sigma))$. Similarly, we have the normalization of the pulse shape, which is normalized in $z$-space. 
\begin{equation}
\begin{aligned}
&1=\sum_{n=1}^N|\omega(z_n)|^2, \quad \sum_{n=1}^N\omega(z_n)\exp(ikz_n)=\tilde{\omega}(k),\\
&M=\sum_{m=1}^M|\tilde{\omega}(k_m)|^2=\sum_{m=1}^M  b\exp(-(k_m-k_0)^2/\beta^2)=\int_{k_0-\Delta k/2}^{k_0+\Delta k/2}\frac{dk}{\Delta k/M} b \exp(-(k-k_0)^2/\beta^2)=\int_{k_0-\Delta k/2}^{k_0+\Delta k/2}dk |\tilde{\omega}(k)|^2
\end{aligned}
\end{equation}
which gives the factor $b=\Delta k/(\sqrt{\pi}\beta\,\text{erf}(\Delta k/2\beta))$. 

We first consider the deviation from the uncorrelated case, as bounded in Proposition 1, given by
\begin{equation}
\begin{aligned}
|P(\vec{s},\vec{\delta})-P_0(\vec{s},\vec{\delta})|\leq(2\epsilon)^2\xi(x_{s_1}-x_{s_{2}} )\xi(x_{s_2}-x_{s_{1}} )
\end{aligned}
\end{equation}
To calculate the function $\xi(x_{s_1}-x_{s_2})$ from Eq. \ref{SI_eq:error_fun}, we  find 
\begin{equation}
\|p(k)\|_\infty=\frac{\sqrt{\beta^2+\sigma^2}}{\sqrt{\pi}\beta\sigma\,\text{erf}\left(\Delta k\sqrt{\beta^2+\sigma^2}/(2\beta\sigma)\right)}
\end{equation}
\begin{equation}
\|p'(k)\|_\infty\leq\sqrt{\frac{2}{e\pi}}\frac{\beta^2+\sigma^2}{\beta^2\sigma^2}\frac{1}{\text{erf}\left(\Delta k\sqrt{\beta^2+\sigma^2}/(2\beta\sigma)\right)}
\end{equation}
This, in turn, provides a bound on the function $\xi(x_{s_1} - x_{s_2})$,
\begin{equation}
\begin{aligned}
&\xi(x_{s_1}-x_{s_2})
\leq \frac{2\|p(k)\|_{\infty}}{|\Delta x|}+\frac{\Delta k\|p'(k)\|_{\infty}}{|\Delta x|}\\
&\leq \frac{2}{|\Delta x|}\frac{\sqrt{\beta^2+\sigma^2}}{\sqrt{\pi}\beta\sigma\,\text{erf}\left(\Delta k\sqrt{\beta^2+\sigma^2}/(2\beta\sigma)\right)}+\frac{\Delta k}{|\Delta x|}\sqrt{\frac{2}{e\pi}}\frac{\beta^2+\sigma^2}{\beta^2\sigma^2}\frac{1}{\text{erf}\left(\Delta k\sqrt{\beta^2+\sigma^2}/(2\beta\sigma)\right)}
\end{aligned}
\end{equation}
To gain better insight into the physical meaning of this result, we assume $\beta = \sigma = \Delta k$, meaning that the widths of the two Gaussian envelopes are equal to the bandwidth.
\begin{equation}
\begin{aligned}
&\xi(x_{s_1}-x_{s_2})
\leq \left(\frac{2\sqrt{2}}{\sqrt{\pi}}+\frac{2\sqrt{2}}{\sqrt{e\pi}}\right)\frac{1}{\text{erf}(1/2)|\Delta x|\Delta k}\approx\frac{4.93}{|\Delta x|\Delta k}
\end{aligned}
\end{equation}
Therefore, if the two photons are detected sufficiently far apart, the uncorrelated-case probability $P_0$ serves as a good approximation.

However, if the deviation cannot be neglected based on the bound above, the correction given in Proposition 2 and Eq.~\ref{SI_eq:two_photon_P} for the two-photon case should be applied:
\begin{equation}
\begin{aligned}\label{SI_eq:P_two_photon2}
&P(\vec{s},\vec{\delta})=\frac{\epsilon^2}{4}(g+g^*+2)(-g-g^*+2)\Omega_{s_1s_1}\Omega_{s_2s_2}+\frac{\epsilon^2}{4}(-g+g^*)(g-g^*)\Omega_{s_1s_2}\Omega_{s_2s_1}.
\end{aligned}
\end{equation}
Here, the coherence function $g$ is the unknown parameter we aim to estimate in the imaging problem, while $\epsilon$ is related to the total source intensity and can be readily estimated. In the absence of correlation, only the first term in $P(\vec{s}, \vec{\delta})$ remains. However, for the Gaussian spectrum considered in this example, the second term cannot be neglected. To account for this correction, we simply need to calculate the correction factor $\Omega_{s_1 s_2}$ as
\begin{equation}
\begin{aligned}
\Omega_{s_1s_2}&=\sum_{m=1}^Mf_m\tilde{\omega}_{s_1}^*(k_m)\tilde{\omega}_{s_2}(k_m)=\int_{k_0-\Delta k/2}^{k_0+\Delta k/2}\frac{dk}{\Delta k/M}\frac{a}{M} \exp(-(k-k_0)^2/\sigma^2)b \exp(-(k-k_0)^2/\beta^2)\exp(ik \Delta x)\\
&=\left(\text{erf}\left[\frac{\Delta k\sigma^2+\beta^2(\Delta k-i\Delta x\sigma^2)}{2\beta\sigma\sqrt{\beta^2+\sigma^2}}\right]+\text{erf}\left[\frac{\Delta k\sigma^2+\beta^2(\Delta k+i\Delta x\sigma^2)}{2\beta\sigma\sqrt{\beta^2+\sigma^2}}\right]\right)\frac{\Delta k \exp\left(-\frac{\beta^2\sigma^2\Delta x^2}{4(\beta^2+\sigma^2)}\right)}{2\sqrt{\pi}\sqrt{\beta^2+\sigma^2}\text{erf}(\Delta k/2\beta)\text{erf}(\Delta k/2\sigma)},
\end{aligned}
\end{equation}
where we use the assumption $\tilde{\omega}_s(k) = e^{-ikx_s} \tilde{\omega}(k)$ and define the spatial distance between 
the two detected photons $\Delta x = x_{s_1} - x_{s_2}$. Again, we assume $\beta = \sigma = \Delta k$ to obtain a more concrete understanding. This gives
\begin{equation}
\begin{aligned}
\Omega_{s_1s_2}
&=\left(\text{erf}\left[\frac{2-i\Delta x\Delta k}{2\sqrt{2}}\right]+\text{erf}\left[\frac{2+i\Delta x\Delta k}{2\sqrt{2}}\right]\right)\frac{\exp\left(-\frac{\Delta k^2\Delta x^2}{8}\right)}{2\sqrt{2\pi}(\text{erf}(1/2))^2},
\end{aligned}
\end{equation}
which, in general, is not a negligible factor when $\Delta x \sim 1/\Delta k$. 
Using parameters similar to the estimation made in Ref.~\cite{gottesman2012longer}, with $\lambda = 800\,\text{nm}$ and $\Delta \lambda = 0.1\,\text{nm}$, we estimate $\Delta k \approx 10^3\,\text{m}^{-1}$. Consequently, when the spatial distance between detected photons $\Delta x \gtrsim 1\,\text{mm}$, the above correction becomes significant and must be taken into account.

In practice, detector resolution time $\Delta t$ is finite. Often, the filter bandwidth $\Delta k$ is chosen so that the corresponding pulse duration $1/c\Delta k$ matches the detector resolution, i.e., $\Delta t \sim 1/c\Delta k$, ensuring that each temporal mode is individually resolved. This standard choice ensures that detection resolves temporal modes arising naturally from the Fourier transform of the filtered frequency modes. However, when detecting thermal light with a non-flat spectrum, adjacent temporal modes resolved by such a detector after the bandpass filter may exhibit residual correlations, which can impact detection outcomes.

\subsection{Heterodyne detection}

For sufficiently strong sources, where each temporal mode is likely to contain more than one photon, single-photon detection at each mode becomes inadequate. A better strategy is to perform local heterodyne detection instead. This corresponds to a Gaussian measurement that projects onto the POVM $\frac{1}{\pi^{2N}} \ket{\vec{\mu}, \vec{\nu}} \bra{\vec{\mu}, \vec{\nu}}$. Given the Gaussian nature of this measurement, the probability distribution for obtaining the outcome $\vec{\zeta}$ can be readily calculated:
\begin{equation}
\begin{aligned}
P(\vec{\mu},\vec{\nu})=\frac{1}{\pi^{2N}\det(I+\Lambda)}\exp(-\vec{\zeta}^\dagger (I+\Lambda)^{-1}\vec{\zeta}),
\end{aligned}
\end{equation}
where $[\vec{\zeta}]_{mn} = \sum_{i=1}^{N} [\vec{\zeta}_i]_{mn}$ for $m = 0, 1$ and $n = 1, 2, \cdots, N$, with $[\vec{\zeta}_i]_{0n} = \mu_i \omega_i(z_n)$ and $[\vec{\zeta}_i]_{1n} = \nu_i \omega_i(z_n)$. If $\Lambda$ becomes block diagonal with $2 \times 2$ blocks in the basis of temporal modes $\vec{\omega}_{i0} = [\omega_i(z_1), \omega_i(z_2), \cdots, \omega_i(z_N), 0, 0, \cdots, 0]$ and $\vec{\omega}_{i1} = [0, 0, \cdots, 0, \omega_i(z_1), \omega_i(z_2), \cdots, \omega_i(z_N)]$, then the measurement outcomes for each temporal mode can be treated as independent. Otherwise, when using an estimator to infer the coherence function, one must account for the full probability distribution in order to properly incorporate the correlations between different temporal modes.

\subsection{Intensity interferometer and correlation functions for the stellar thermal sources}

In this subsection, we provide further details on how the intensity interferometer operates in the case of a general spectrum. In such a setup, photon number measurements are performed locally at each telescope across all localized temporal modes. The corresponding POVM elements are denoted by $\{\ket{\vec{p},\vec{s}}\}_{\vec{p},\vec{s}}$, where $\vec{p} \in \{0,1\}^{\otimes Q}$ and $\vec{s} \in \{1,2,\dots,N\}^{\otimes Q}$, with $Q = 0,1,2,\dots,\infty$. Here, $s_i$ labels the temporal mode in which the $i$th photon is detected, and $p_i \in \{0,1\}$ denotes the telescope (e.g., 0 for the first telescope and 1 for the second) that registers the detection. Using this measurement model, we can compute the associated probability distribution
\begin{equation}
\begin{aligned}
P(\vec{s},\vec{p})=\bra{\vec{s},\vec{p}}\rho\ket{\vec{s},\vec{p}}=\epsilon^Q A^2(\vec{s},\vec{p})\sum_\mathbb{P}\prod_{i=1}^Q(\Omega_{s_is_{\mathbb{P}_i}}g(p_i,p_{\mathbb{P}_i})),
\end{aligned}
\end{equation}
where we have used Eq.~\ref{SI_eq:g_pq}. 

As a simple example, let us consider the case of $Q=2$:
\begin{equation}
\begin{aligned}\label{SI_eq:P_HBT}
P(\vec{s},\vec{p})=\epsilon^2A^2(\vec{s},\vec{p})(\Omega_{s_1s_1}\Omega_{s_2s_2}g(p_1,p_1)g(p_2,p_2)+\Omega_{s_1s_2}\Omega_{s_2s_1}g(p_1,p_2)g(p_2,p_1)).
\end{aligned}
\end{equation}
If a thermal source has a flat spectrum, 
\begin{equation}
\begin{aligned}
P(\vec{s},\vec{p})=\left\{ 
  \begin{array}{ll}
   \epsilon^2\Omega_{ss}^2\quad &\text{if}\quad s_1\neq s_2\\
   \epsilon^2\Omega_{ss}^2\quad &\text{if}\quad s_1= s_2,\,p_1=p_2\\
   \epsilon^2\Omega_{ss}^2(1+|g|^2)\quad &\text{if}\quad s_1= s_2,\,p_1\neq p_2\\
  \end{array}
\right..
\end{aligned}
\end{equation}
Therefore, when the two detected photons originate from different temporal modes, i.e., $s_1 \neq s_2$, no information about the coherence function $g$ can be extracted. For the intensity interferometer to be sensitive to $g$, it is necessary that both photons be detected in the same time but at different telescopes, that is, $s_1 = s_2$ and $p_1 \neq p_2$.

If a thermal source has a non-flat spectrum
\begin{equation}
\begin{aligned}
P(\vec{s},\vec{p})=\left\{ 
  \begin{array}{ll}
   \epsilon^2(\Omega_{ss}^2+\Omega_{s_1s_2}\Omega_{s_2s_1})\quad &\text{if}\quad s_1\neq s_2,\,p_1=p_2\\
   \epsilon^2(\Omega_{ss}^2+\Omega_{s_1s_2}\Omega_{s_2s_1}|g|^2)\quad &\text{if}\quad s_1\neq s_2,\,p_1\neq p_2\\
   \epsilon^2\Omega_{ss}^2\quad &\text{if}\quad s_1= s_2,\,p_1=p_2\\
   \epsilon^2\Omega_{ss}^2(1+|g|^2)\quad &\text{if}\quad s_1= s_2,\,p_1\neq p_2\\
  \end{array}
\right.
\end{aligned}
\end{equation}
The inter-mode correlations allow us to extract some information about $|g|$ even when the two photons are detected at different times, i.e., $s_1 \neq s_2$. Nevertheless, it is important to note that the overlap function $\Omega_{s_1 s_2}$ decays as the temporal separation between modes increases.

The operation of an intensity interferometer is typically analyzed in terms of the second-order correlation function. Therefore, we now present the  expression for the correlation functions relevant to our setup.
\begin{equation}
\begin{aligned}
&G^{(1)}(s_1p_1;s_2p_2)=\tr(\rho a_{s_1p_1}^\dagger a_{s_2p_2})=\vec{\omega}_{s_1p_1}^\dagger\Lambda\vec{\omega}_{s_2p_2}=\epsilon\Omega_{s_1s_2}g(p_1,p_2)\\
&G^{(2)}(s_1p_1,s_2p_2;s_2p_2,s_1p_1)=\tr(\rho a_{s_1p_1}^\dagger a_{s_2p_2}^\dagger a_{s_2p_2}a_{s_1p_1})=(\vec{\omega}_{s_1p_1}^\dagger\Lambda\vec{\omega}_{s_1p_1})(\vec{\omega}_{s_2p_2}^\dagger\Lambda\vec{\omega}_{s_2p_2})+(\vec{\omega}_{s_1p_1}^\dagger\Lambda\vec{\omega}_{s_2p_2})(\vec{\omega}_{s_2p_2}^\dagger\Lambda\vec{\omega}_{s_1p_1})\\
&=\epsilon^2\Omega_{s_1s_1}\Omega_{s_2s_2}g(p_1,p_1)g(p_2,p_2)+\epsilon^2\Omega_{s_1s_2}\Omega_{s_2s_1}g(p_1,p_2)g(p_2,p_1).
\end{aligned}
\end{equation}
The second-order correlation function closely resembles the probability distribution given in Eq.~\ref{SI_eq:P_HBT}. In particular, quantum statistical effects arising from the thermal nature of the light can be observed by selecting events where $s_1 = s_2 = s$, i.e., when both photons are detected at the same time:
\begin{equation}
G^{(2)}(sp_1,sp_2;sp_2,sp_1)/(G^{(1)}(sp_1;sp_2))^2=(\Omega_{ss}^2+\Omega_{ss}^2g(p_1,p_2)g(p_2,p_1))/\Omega_{ss}^2 g^2(p_1,p_2)
\end{equation}
If we further choose $p_1=p_2=p$, 
\begin{equation}
G^{(2)}(sp,sp;sp,sp)/(G^{(1)} (sp;sp))^2=(\Omega_{ss}^2+\Omega_{ss}^2)/\Omega_{ss}^2 =2
\end{equation}
which is as expected for thermal light.

\bibliography{arxiv}

\begin{thebibliography}{26}%
\makeatletter
\providecommand \@ifxundefined [1]{%
 \@ifx{#1\undefined}
}%
\providecommand \@ifnum [1]{%
 \ifnum #1\expandafter \@firstoftwo
 \else \expandafter \@secondoftwo
 \fi
}%
\providecommand \@ifx [1]{%
 \ifx #1\expandafter \@firstoftwo
 \else \expandafter \@secondoftwo
 \fi
}%
\providecommand \natexlab [1]{#1}%
\providecommand \enquote  [1]{``#1''}%
\providecommand \bibnamefont  [1]{#1}%
\providecommand \bibfnamefont [1]{#1}%
\providecommand \citenamefont [1]{#1}%
\providecommand \href@noop [0]{\@secondoftwo}%
\providecommand \href [0]{\begingroup \@sanitize@url \@href}%
\providecommand \@href[1]{\@@startlink{#1}\@@href}%
\providecommand \@@href[1]{\endgroup#1\@@endlink}%
\providecommand \@sanitize@url [0]{\catcode `\\12\catcode `\$12\catcode
  `\&12\catcode `\#12\catcode `\^12\catcode `\_12\catcode `\%12\relax}%
\providecommand \@@startlink[1]{}%
\providecommand \@@endlink[0]{}%
\providecommand \url  [0]{\begingroup\@sanitize@url \@url }%
\providecommand \@url [1]{\endgroup\@href {#1}{\urlprefix }}%
\providecommand \urlprefix  [0]{URL }%
\providecommand \Eprint [0]{\href }%
\providecommand \doibase [0]{https://doi.org/}%
\providecommand \selectlanguage [0]{\@gobble}%
\providecommand \bibinfo  [0]{\@secondoftwo}%
\providecommand \bibfield  [0]{\@secondoftwo}%
\providecommand \translation [1]{[#1]}%
\providecommand \BibitemOpen [0]{}%
\providecommand \bibitemStop [0]{}%
\providecommand \bibitemNoStop [0]{.\EOS\space}%
\providecommand \EOS [0]{\spacefactor3000\relax}%
\providecommand \BibitemShut  [1]{\csname bibitem#1\endcsname}%
\let\auto@bib@innerbib\@empty
\bibitem [{\citenamefont {Mandel}\ and\ \citenamefont
  {Wolf}(1995)}]{mandel1995optical}%
  \BibitemOpen
  \bibfield  {author} {\bibinfo {author} {\bibfnamefont {L.}~\bibnamefont
  {Mandel}}\ and\ \bibinfo {author} {\bibfnamefont {E.}~\bibnamefont {Wolf}},\
  }\href@noop {} {\emph {\bibinfo {title} {Optical coherence and quantum
  optics}}}\ (\bibinfo  {publisher} {Cambridge University Press},\ \bibinfo
  {year} {1995})\BibitemShut {NoStop}%
\bibitem [{\citenamefont {Gottesman}\ \emph {et~al.}(2012)\citenamefont
  {Gottesman}, \citenamefont {Jennewein},\ and\ \citenamefont
  {Croke}}]{gottesman2012longer}%
  \BibitemOpen
  \bibfield  {author} {\bibinfo {author} {\bibfnamefont {D.}~\bibnamefont
  {Gottesman}}, \bibinfo {author} {\bibfnamefont {T.}~\bibnamefont
  {Jennewein}},\ and\ \bibinfo {author} {\bibfnamefont {S.}~\bibnamefont
  {Croke}},\ }\bibfield  {title} {\bibinfo {title} {Longer-baseline telescopes
  using quantum repeaters},\ }\href@noop {} {\bibfield  {journal} {\bibinfo
  {journal} {Physical review letters}\ }\textbf {\bibinfo {volume} {109}},\
  \bibinfo {pages} {070503} (\bibinfo {year} {2012})}\BibitemShut {NoStop}%
\bibitem [{\citenamefont {Duan}\ and\ \citenamefont
  {Monroe}(2010)}]{duan2010colloquium}%
  \BibitemOpen
  \bibfield  {author} {\bibinfo {author} {\bibfnamefont {L.-M.}\ \bibnamefont
  {Duan}}\ and\ \bibinfo {author} {\bibfnamefont {C.}~\bibnamefont {Monroe}},\
  }\bibfield  {title} {\bibinfo {title} {Colloquium: Quantum networks with
  trapped ions},\ }\href@noop {} {\bibfield  {journal} {\bibinfo  {journal}
  {Reviews of Modern Physics}\ }\textbf {\bibinfo {volume} {82}},\ \bibinfo
  {pages} {1209} (\bibinfo {year} {2010})}\BibitemShut {NoStop}%
\bibitem [{\citenamefont {Wei}\ \emph {et~al.}(2022)\citenamefont {Wei},
  \citenamefont {Jing}, \citenamefont {Zhang}, \citenamefont {Liao},
  \citenamefont {Yuan}, \citenamefont {Fan}, \citenamefont {Lyu}, \citenamefont
  {Zhou}, \citenamefont {Wang}, \citenamefont {Deng} \emph
  {et~al.}}]{wei2022towards}%
  \BibitemOpen
  \bibfield  {author} {\bibinfo {author} {\bibfnamefont {S.-H.}\ \bibnamefont
  {Wei}}, \bibinfo {author} {\bibfnamefont {B.}~\bibnamefont {Jing}}, \bibinfo
  {author} {\bibfnamefont {X.-Y.}\ \bibnamefont {Zhang}}, \bibinfo {author}
  {\bibfnamefont {J.-Y.}\ \bibnamefont {Liao}}, \bibinfo {author}
  {\bibfnamefont {C.-Z.}\ \bibnamefont {Yuan}}, \bibinfo {author}
  {\bibfnamefont {B.-Y.}\ \bibnamefont {Fan}}, \bibinfo {author} {\bibfnamefont
  {C.}~\bibnamefont {Lyu}}, \bibinfo {author} {\bibfnamefont {D.-L.}\
  \bibnamefont {Zhou}}, \bibinfo {author} {\bibfnamefont {Y.}~\bibnamefont
  {Wang}}, \bibinfo {author} {\bibfnamefont {G.-W.}\ \bibnamefont {Deng}},
  \emph {et~al.},\ }\bibfield  {title} {\bibinfo {title} {Towards real-world
  quantum networks: a review},\ }\href@noop {} {\bibfield  {journal} {\bibinfo
  {journal} {Laser \& Photonics Reviews}\ }\textbf {\bibinfo {volume} {16}},\
  \bibinfo {pages} {2100219} (\bibinfo {year} {2022})}\BibitemShut {NoStop}%
\bibitem [{\citenamefont {Sangouard}\ \emph {et~al.}(2011)\citenamefont
  {Sangouard}, \citenamefont {Simon}, \citenamefont {De~Riedmatten},\ and\
  \citenamefont {Gisin}}]{sangouard2011quantum}%
  \BibitemOpen
  \bibfield  {author} {\bibinfo {author} {\bibfnamefont {N.}~\bibnamefont
  {Sangouard}}, \bibinfo {author} {\bibfnamefont {C.}~\bibnamefont {Simon}},
  \bibinfo {author} {\bibfnamefont {H.}~\bibnamefont {De~Riedmatten}},\ and\
  \bibinfo {author} {\bibfnamefont {N.}~\bibnamefont {Gisin}},\ }\bibfield
  {title} {\bibinfo {title} {Quantum repeaters based on atomic ensembles and
  linear optics},\ }\href@noop {} {\bibfield  {journal} {\bibinfo  {journal}
  {Reviews of Modern Physics}\ }\textbf {\bibinfo {volume} {83}},\ \bibinfo
  {pages} {33} (\bibinfo {year} {2011})}\BibitemShut {NoStop}%
\bibitem [{\citenamefont {Azuma}\ \emph {et~al.}(2023)\citenamefont {Azuma},
  \citenamefont {Economou}, \citenamefont {Elkouss}, \citenamefont {Hilaire},
  \citenamefont {Jiang}, \citenamefont {Lo},\ and\ \citenamefont
  {Tzitrin}}]{azuma2023quantum}%
  \BibitemOpen
  \bibfield  {author} {\bibinfo {author} {\bibfnamefont {K.}~\bibnamefont
  {Azuma}}, \bibinfo {author} {\bibfnamefont {S.~E.}\ \bibnamefont {Economou}},
  \bibinfo {author} {\bibfnamefont {D.}~\bibnamefont {Elkouss}}, \bibinfo
  {author} {\bibfnamefont {P.}~\bibnamefont {Hilaire}}, \bibinfo {author}
  {\bibfnamefont {L.}~\bibnamefont {Jiang}}, \bibinfo {author} {\bibfnamefont
  {H.-K.}\ \bibnamefont {Lo}},\ and\ \bibinfo {author} {\bibfnamefont
  {I.}~\bibnamefont {Tzitrin}},\ }\bibfield  {title} {\bibinfo {title} {Quantum
  repeaters: From quantum networks to the quantum internet},\ }\href@noop {}
  {\bibfield  {journal} {\bibinfo  {journal} {Reviews of Modern Physics}\
  }\textbf {\bibinfo {volume} {95}},\ \bibinfo {pages} {045006} (\bibinfo
  {year} {2023})}\BibitemShut {NoStop}%
\bibitem [{\citenamefont {Khabiboulline}\ \emph
  {et~al.}(2019{\natexlab{a}})\citenamefont {Khabiboulline}, \citenamefont
  {Borregaard}, \citenamefont {De~Greve},\ and\ \citenamefont
  {Lukin}}]{khabiboulline2019quantum}%
  \BibitemOpen
  \bibfield  {author} {\bibinfo {author} {\bibfnamefont {E.~T.}\ \bibnamefont
  {Khabiboulline}}, \bibinfo {author} {\bibfnamefont {J.}~\bibnamefont
  {Borregaard}}, \bibinfo {author} {\bibfnamefont {K.}~\bibnamefont
  {De~Greve}},\ and\ \bibinfo {author} {\bibfnamefont {M.~D.}\ \bibnamefont
  {Lukin}},\ }\bibfield  {title} {\bibinfo {title} {Quantum-assisted telescope
  arrays},\ }\href@noop {} {\bibfield  {journal} {\bibinfo  {journal} {Physical
  review A}\ }\textbf {\bibinfo {volume} {100}},\ \bibinfo {pages} {022316}
  (\bibinfo {year} {2019}{\natexlab{a}})}\BibitemShut {NoStop}%
\bibitem [{\citenamefont {Khabiboulline}\ \emph
  {et~al.}(2019{\natexlab{b}})\citenamefont {Khabiboulline}, \citenamefont
  {Borregaard}, \citenamefont {De~Greve},\ and\ \citenamefont
  {Lukin}}]{khabiboulline2019optical}%
  \BibitemOpen
  \bibfield  {author} {\bibinfo {author} {\bibfnamefont {E.~T.}\ \bibnamefont
  {Khabiboulline}}, \bibinfo {author} {\bibfnamefont {J.}~\bibnamefont
  {Borregaard}}, \bibinfo {author} {\bibfnamefont {K.}~\bibnamefont
  {De~Greve}},\ and\ \bibinfo {author} {\bibfnamefont {M.~D.}\ \bibnamefont
  {Lukin}},\ }\bibfield  {title} {\bibinfo {title} {Optical interferometry with
  quantum networks},\ }\href@noop {} {\bibfield  {journal} {\bibinfo  {journal}
  {Physical review letters}\ }\textbf {\bibinfo {volume} {123}},\ \bibinfo
  {pages} {070504} (\bibinfo {year} {2019}{\natexlab{b}})}\BibitemShut
  {NoStop}%
\bibitem [{\citenamefont {Czupryniak}\ \emph {et~al.}(2023)\citenamefont
  {Czupryniak}, \citenamefont {Steinmetz}, \citenamefont {Kwiat},\ and\
  \citenamefont {Jordan}}]{czupryniak2023optimal}%
  \BibitemOpen
  \bibfield  {author} {\bibinfo {author} {\bibfnamefont {R.}~\bibnamefont
  {Czupryniak}}, \bibinfo {author} {\bibfnamefont {J.}~\bibnamefont
  {Steinmetz}}, \bibinfo {author} {\bibfnamefont {P.~G.}\ \bibnamefont
  {Kwiat}},\ and\ \bibinfo {author} {\bibfnamefont {A.~N.}\ \bibnamefont
  {Jordan}},\ }\bibfield  {title} {\bibinfo {title} {Optimal qubit circuits for
  quantum-enhanced telescopes},\ }\href@noop {} {\bibfield  {journal} {\bibinfo
   {journal} {Physical Review A}\ }\textbf {\bibinfo {volume} {108}},\ \bibinfo
  {pages} {052408} (\bibinfo {year} {2023})}\BibitemShut {NoStop}%
\bibitem [{\citenamefont {Czupryniak}\ \emph {et~al.}(2022)\citenamefont
  {Czupryniak}, \citenamefont {Chitambar}, \citenamefont {Steinmetz},\ and\
  \citenamefont {Jordan}}]{czupryniak2022quantum}%
  \BibitemOpen
  \bibfield  {author} {\bibinfo {author} {\bibfnamefont {R.}~\bibnamefont
  {Czupryniak}}, \bibinfo {author} {\bibfnamefont {E.}~\bibnamefont
  {Chitambar}}, \bibinfo {author} {\bibfnamefont {J.}~\bibnamefont
  {Steinmetz}},\ and\ \bibinfo {author} {\bibfnamefont {A.~N.}\ \bibnamefont
  {Jordan}},\ }\bibfield  {title} {\bibinfo {title} {Quantum telescopy clock
  games},\ }\href@noop {} {\bibfield  {journal} {\bibinfo  {journal} {Physical
  Review A}\ }\textbf {\bibinfo {volume} {106}},\ \bibinfo {pages} {032424}
  (\bibinfo {year} {2022})}\BibitemShut {NoStop}%
\bibitem [{\citenamefont {Marchese}\ and\ \citenamefont
  {Kok}(2023)}]{marchese2023large}%
  \BibitemOpen
  \bibfield  {author} {\bibinfo {author} {\bibfnamefont {M.~M.}\ \bibnamefont
  {Marchese}}\ and\ \bibinfo {author} {\bibfnamefont {P.}~\bibnamefont {Kok}},\
  }\bibfield  {title} {\bibinfo {title} {Large baseline optical imaging
  assisted by single photons and linear quantum optics},\ }\href@noop {}
  {\bibfield  {journal} {\bibinfo  {journal} {Physical Review Letters}\
  }\textbf {\bibinfo {volume} {130}},\ \bibinfo {pages} {160801} (\bibinfo
  {year} {2023})}\BibitemShut {NoStop}%
\bibitem [{\citenamefont {Huang}\ \emph {et~al.}(2024)\citenamefont {Huang},
  \citenamefont {Baragiola}, \citenamefont {Menicucci},\ and\ \citenamefont
  {Wilde}}]{huang2024limited}%
  \BibitemOpen
  \bibfield  {author} {\bibinfo {author} {\bibfnamefont {Z.}~\bibnamefont
  {Huang}}, \bibinfo {author} {\bibfnamefont {B.~Q.}\ \bibnamefont
  {Baragiola}}, \bibinfo {author} {\bibfnamefont {N.~C.}\ \bibnamefont
  {Menicucci}},\ and\ \bibinfo {author} {\bibfnamefont {M.~M.}\ \bibnamefont
  {Wilde}},\ }\bibfield  {title} {\bibinfo {title} {Limited quantum advantage
  for stellar interferometry via continuous-variable teleportation},\
  }\href@noop {} {\bibfield  {journal} {\bibinfo  {journal} {Physical Review
  A}\ }\textbf {\bibinfo {volume} {109}},\ \bibinfo {pages} {052434} (\bibinfo
  {year} {2024})}\BibitemShut {NoStop}%
\bibitem [{\citenamefont {Wang}\ \emph {et~al.}(2025)\citenamefont {Wang},
  \citenamefont {Zhang},\ and\ \citenamefont {Lorenz}}]{wang2023astronomical}%
  \BibitemOpen
  \bibfield  {author} {\bibinfo {author} {\bibfnamefont {Y.}~\bibnamefont
  {Wang}}, \bibinfo {author} {\bibfnamefont {Y.}~\bibnamefont {Zhang}},\ and\
  \bibinfo {author} {\bibfnamefont {V.~O.}\ \bibnamefont {Lorenz}},\ }\bibfield
   {title} {\bibinfo {title} {Astronomical interferometry using continuous
  variable quantum teleportation},\ }\href@noop {} {\bibfield  {journal}
  {\bibinfo  {journal} {Physical Review Research}\ }\textbf {\bibinfo {volume}
  {7}},\ \bibinfo {pages} {023154} (\bibinfo {year} {2025})}\BibitemShut
  {NoStop}%
\bibitem [{\citenamefont {Purvis}\ \emph {et~al.}(2024)\citenamefont {Purvis},
  \citenamefont {Lafler},\ and\ \citenamefont {Lanning}}]{purvis2024practical}%
  \BibitemOpen
  \bibfield  {author} {\bibinfo {author} {\bibfnamefont {B.}~\bibnamefont
  {Purvis}}, \bibinfo {author} {\bibfnamefont {R.}~\bibnamefont {Lafler}},\
  and\ \bibinfo {author} {\bibfnamefont {R.~N.}\ \bibnamefont {Lanning}},\
  }\bibfield  {title} {\bibinfo {title} {Practical approach to extending
  baselines of telescopes using continuous-variable quantum information},\
  }\href@noop {} {\bibfield  {journal} {\bibinfo  {journal} {New Journal of
  Physics}\ }\textbf {\bibinfo {volume} {26}},\ \bibinfo {pages} {103006}
  (\bibinfo {year} {2024})}\BibitemShut {NoStop}%
\bibitem [{\citenamefont {Zhang}\ and\ \citenamefont
  {Jennewein}(2025)}]{zhang2025criteria}%
  \BibitemOpen
  \bibfield  {author} {\bibinfo {author} {\bibfnamefont {Y.}~\bibnamefont
  {Zhang}}\ and\ \bibinfo {author} {\bibfnamefont {T.}~\bibnamefont
  {Jennewein}},\ }\bibfield  {title} {\bibinfo {title} {Criteria for optimal
  entanglement-assisted long baseline imaging protocols},\ }\href@noop {}
  {\bibfield  {journal} {\bibinfo  {journal} {arXiv preprint arXiv:2501.16670}\
  } (\bibinfo {year} {2025})}\BibitemShut {NoStop}%
\bibitem [{\citenamefont {Chenu}\ \emph
  {et~al.}(2015{\natexlab{a}})\citenamefont {Chenu}, \citenamefont
  {Bra{\'n}czyk},\ and\ \citenamefont {Sipe}}]{chenu2015first}%
  \BibitemOpen
  \bibfield  {author} {\bibinfo {author} {\bibfnamefont {A.}~\bibnamefont
  {Chenu}}, \bibinfo {author} {\bibfnamefont {A.~M.}\ \bibnamefont
  {Bra{\'n}czyk}},\ and\ \bibinfo {author} {\bibfnamefont {J.}~\bibnamefont
  {Sipe}},\ }\bibfield  {title} {\bibinfo {title} {First-order decomposition of
  thermal light in terms of a statistical mixture of single pulses},\
  }\href@noop {} {\bibfield  {journal} {\bibinfo  {journal} {Physical Review
  A}\ }\textbf {\bibinfo {volume} {91}},\ \bibinfo {pages} {063813} (\bibinfo
  {year} {2015}{\natexlab{a}})}\BibitemShut {NoStop}%
\bibitem [{\citenamefont {Chenu}\ \emph
  {et~al.}(2015{\natexlab{b}})\citenamefont {Chenu}, \citenamefont
  {Bra{\'n}czyk}, \citenamefont {Scholes},\ and\ \citenamefont
  {Sipe}}]{chenu2015thermal}%
  \BibitemOpen
  \bibfield  {author} {\bibinfo {author} {\bibfnamefont {A.}~\bibnamefont
  {Chenu}}, \bibinfo {author} {\bibfnamefont {A.~M.}\ \bibnamefont
  {Bra{\'n}czyk}}, \bibinfo {author} {\bibfnamefont {G.~D.}\ \bibnamefont
  {Scholes}},\ and\ \bibinfo {author} {\bibfnamefont {J.~E.}\ \bibnamefont
  {Sipe}},\ }\bibfield  {title} {\bibinfo {title} {Thermal light cannot be
  represented as a statistical mixture of single pulses},\ }\href@noop {}
  {\bibfield  {journal} {\bibinfo  {journal} {Physical review letters}\
  }\textbf {\bibinfo {volume} {114}},\ \bibinfo {pages} {213601} (\bibinfo
  {year} {2015}{\natexlab{b}})}\BibitemShut {NoStop}%
\bibitem [{\citenamefont {Bra{\'n}czyk}\ \emph {et~al.}(2017)\citenamefont
  {Bra{\'n}czyk}, \citenamefont {Chenu},\ and\ \citenamefont
  {Sipe}}]{branczyk2017thermal}%
  \BibitemOpen
  \bibfield  {author} {\bibinfo {author} {\bibfnamefont {A.~M.}\ \bibnamefont
  {Bra{\'n}czyk}}, \bibinfo {author} {\bibfnamefont {A.}~\bibnamefont
  {Chenu}},\ and\ \bibinfo {author} {\bibfnamefont {J.}~\bibnamefont {Sipe}},\
  }\bibfield  {title} {\bibinfo {title} {Thermal light as a mixture of sets of
  pulses: the quasi-1d example},\ }\href@noop {} {\bibfield  {journal}
  {\bibinfo  {journal} {JOSA B}\ }\textbf {\bibinfo {volume} {34}},\ \bibinfo
  {pages} {1536} (\bibinfo {year} {2017})}\BibitemShut {NoStop}%
\bibitem [{\citenamefont {Glauber}(1963)}]{glauber1963quantum}%
  \BibitemOpen
  \bibfield  {author} {\bibinfo {author} {\bibfnamefont {R.~J.}\ \bibnamefont
  {Glauber}},\ }\bibfield  {title} {\bibinfo {title} {The quantum theory of
  optical coherence},\ }\href@noop {} {\bibfield  {journal} {\bibinfo
  {journal} {Physical Review}\ }\textbf {\bibinfo {volume} {130}},\ \bibinfo
  {pages} {2529} (\bibinfo {year} {1963})}\BibitemShut {NoStop}%
\bibitem [{\citenamefont {Brown}\ and\ \citenamefont
  {Twiss}(1956)}]{brown1956correlation}%
  \BibitemOpen
  \bibfield  {author} {\bibinfo {author} {\bibfnamefont {R.~H.}\ \bibnamefont
  {Brown}}\ and\ \bibinfo {author} {\bibfnamefont {R.~Q.}\ \bibnamefont
  {Twiss}},\ }\bibfield  {title} {\bibinfo {title} {Correlation between photons
  in two coherent beams of light},\ }\href@noop {} {\bibfield  {journal}
  {\bibinfo  {journal} {Nature}\ }\textbf {\bibinfo {volume} {177}},\ \bibinfo
  {pages} {27} (\bibinfo {year} {1956})}\BibitemShut {NoStop}%
\bibitem [{\citenamefont {Hanbury~Brown}\ and\ \citenamefont
  {Twiss}(1979)}]{hanbury1979test}%
  \BibitemOpen
  \bibfield  {author} {\bibinfo {author} {\bibfnamefont {R.}~\bibnamefont
  {Hanbury~Brown}}\ and\ \bibinfo {author} {\bibfnamefont {R.~Q.}\ \bibnamefont
  {Twiss}},\ }\bibfield  {title} {\bibinfo {title} {A test of a new type of
  stellar interferometer on sirius},\ }in\ \href@noop {} {\emph {\bibinfo
  {booktitle} {A Source Book in Astronomy and Astrophysics, 1900--1975}}}\
  (\bibinfo  {publisher} {Harvard University Press},\ \bibinfo {year} {1979})\
  pp.\ \bibinfo {pages} {8--12}\BibitemShut {NoStop}%
\bibitem [{\citenamefont {Monnier}(2003)}]{monnier2003optical}%
  \BibitemOpen
  \bibfield  {author} {\bibinfo {author} {\bibfnamefont {J.~D.}\ \bibnamefont
  {Monnier}},\ }\bibfield  {title} {\bibinfo {title} {Optical interferometry in
  astronomy},\ }\href@noop {} {\bibfield  {journal} {\bibinfo  {journal}
  {Reports on Progress in Physics}\ }\textbf {\bibinfo {volume} {66}},\
  \bibinfo {pages} {789} (\bibinfo {year} {2003})}\BibitemShut {NoStop}%
\bibitem [{\citenamefont {Blow}\ \emph {et~al.}(1990)\citenamefont {Blow},
  \citenamefont {Loudon}, \citenamefont {Phoenix},\ and\ \citenamefont
  {Shepherd}}]{blow1990continuum}%
  \BibitemOpen
  \bibfield  {author} {\bibinfo {author} {\bibfnamefont {K.}~\bibnamefont
  {Blow}}, \bibinfo {author} {\bibfnamefont {R.}~\bibnamefont {Loudon}},
  \bibinfo {author} {\bibfnamefont {S.~J.}\ \bibnamefont {Phoenix}},\ and\
  \bibinfo {author} {\bibfnamefont {T.}~\bibnamefont {Shepherd}},\ }\bibfield
  {title} {\bibinfo {title} {Continuum fields in quantum optics},\ }\href@noop
  {} {\bibfield  {journal} {\bibinfo  {journal} {Physical Review A}\ }\textbf
  {\bibinfo {volume} {42}},\ \bibinfo {pages} {4102} (\bibinfo {year}
  {1990})}\BibitemShut {NoStop}%
\bibitem [{\citenamefont {Rohde}\ \emph {et~al.}(2007)\citenamefont {Rohde},
  \citenamefont {Mauerer},\ and\ \citenamefont
  {Silberhorn}}]{rohde2007spectral}%
  \BibitemOpen
  \bibfield  {author} {\bibinfo {author} {\bibfnamefont {P.~P.}\ \bibnamefont
  {Rohde}}, \bibinfo {author} {\bibfnamefont {W.}~\bibnamefont {Mauerer}},\
  and\ \bibinfo {author} {\bibfnamefont {C.}~\bibnamefont {Silberhorn}},\
  }\bibfield  {title} {\bibinfo {title} {Spectral structure and decompositions
  of optical states, and their applications},\ }\href@noop {} {\bibfield
  {journal} {\bibinfo  {journal} {New Journal of Physics}\ }\textbf {\bibinfo
  {volume} {9}},\ \bibinfo {pages} {91} (\bibinfo {year} {2007})}\BibitemShut
  {NoStop}%
\bibitem [{\citenamefont {Weedbrook}\ \emph {et~al.}(2012)\citenamefont
  {Weedbrook}, \citenamefont {Pirandola}, \citenamefont
  {Garc{\'\i}a-Patr{\'o}n}, \citenamefont {Cerf}, \citenamefont {Ralph},
  \citenamefont {Shapiro},\ and\ \citenamefont
  {Lloyd}}]{weedbrook2012gaussian}%
  \BibitemOpen
  \bibfield  {author} {\bibinfo {author} {\bibfnamefont {C.}~\bibnamefont
  {Weedbrook}}, \bibinfo {author} {\bibfnamefont {S.}~\bibnamefont
  {Pirandola}}, \bibinfo {author} {\bibfnamefont {R.}~\bibnamefont
  {Garc{\'\i}a-Patr{\'o}n}}, \bibinfo {author} {\bibfnamefont {N.~J.}\
  \bibnamefont {Cerf}}, \bibinfo {author} {\bibfnamefont {T.~C.}\ \bibnamefont
  {Ralph}}, \bibinfo {author} {\bibfnamefont {J.~H.}\ \bibnamefont {Shapiro}},\
  and\ \bibinfo {author} {\bibfnamefont {S.}~\bibnamefont {Lloyd}},\ }\bibfield
   {title} {\bibinfo {title} {Gaussian quantum information},\ }\href@noop {}
  {\bibfield  {journal} {\bibinfo  {journal} {Reviews of Modern Physics}\
  }\textbf {\bibinfo {volume} {84}},\ \bibinfo {pages} {621} (\bibinfo {year}
  {2012})}\BibitemShut {NoStop}%
\bibitem [{\citenamefont {Leonhardt}(2003)}]{leonhardt2003quantum}%
  \BibitemOpen
  \bibfield  {author} {\bibinfo {author} {\bibfnamefont {U.}~\bibnamefont
  {Leonhardt}},\ }\bibfield  {title} {\bibinfo {title} {Quantum physics of
  simple optical instruments},\ }\href@noop {} {\bibfield  {journal} {\bibinfo
  {journal} {Reports on Progress in Physics}\ }\textbf {\bibinfo {volume}
  {66}},\ \bibinfo {pages} {1207} (\bibinfo {year} {2003})}\BibitemShut
  {NoStop}%
\end{thebibliography}%

\end{document}